\documentclass[a4paper,fleqn,usenatbib]{mnras}
\usepackage[T1]{fontenc}
\usepackage{ae,aecompl}
\usepackage{graphicx}
\usepackage{amsmath}
\usepackage{amssymb}

\newcommand{\mkm}{$\mu$m}
\newcommand{\kms}{\,km\,s$^{-1}$}

\usepackage{longtable}

\title[High-resolution spectroscopy] {High-resolution spectroscopy
of the high velocity hot post-AGB star IRAS~18379--1707 (LS~5112)}
\author[Ikonnikova et al.]{N.P. Ikonnikova$^{1}$,  M.
Parthasarathy$^{2}$\thanks{E-mail:m-partha@hotmail.com}, A.V. Dodin$^{1}$, S. Hubrig$^{4}$, G. Sarkar$^{5}$\\
$^{1}$ Lomonosov Moscow State University, Sternberg Astronomical
Institute, 13 Universitetskij prospekt, Moscow 119234, Russia\\
$^{2}$ Indian Institute of Astrophysics, Bangalore 560034, India\\
$^{4}$ Leibniz Institute for Astrophysics (AIP), Potsdam, D-14482,
Germany\\
$^{5}$ Department of Physics, Indian Institute of Technology,
Kanpur-208016, UP, India}

\date{Accepted 2019. Received 2019; in original form 2019}

\pubyear{2019}

\begin{document}
\label{firstpage}
\pagerange{\pageref{firstpage}--\pageref{lastpage}} \maketitle

\begin{abstract}

The high-resolution ($R\sim48\,000$) optical spectrum of the
B-type supergiant LS~5112, identified as the optical counterpart
of the post-AGB candidate IRAS~18379--1707 is analysed. We report
the detailed identifications of the observed absorption and
emission features in the wavelength range 3700-9200 \AA\ for the
first time. The absorption line spectrum has been analysed using
non-LTE model atmosphere techniques to determine stellar
atmospheric parameters and chemical composition. We estimate
$T_\text{eff}=18~000\pm1000$ K, $\log g=2.25\pm0.08$, $\xi_{\rm
t}=10\pm4$\,{\kms} and $v \sin i=37\pm6$ {\kms}, and the derived
abundances indicate a metal-deficient ([M/H]$\approx-0.6)$
post-AGB star. Chemical abundances of eight different elements
were obtained. The estimates of the CNO abundances in IRAS
18379--1707 indicate that these elements are overabundant with
[(C+N+O)/S]=+0.5$\pm$0.2 suggesting that the products of helium
burning have been brought to the surface as a result of third
dredge-up on the AGB. From the absorption lines, we derived
heliocentric radial velocity of $V_\text{r}=-124.0\pm0.4$
{\kms}. We have identified permitted emission lines of \ion{O}{I},
\ion{N}{I}, \ion{Na}{I}, \ion{S}{II}, \ion{Si}{II}, \ion{C}{II},
\ion{Mg}{II} and \ion{Fe}{III}. The nebula forbidden lines of
[\ion{N}{I}], [\ion{O}{I}], [\ion{Fe}{II}], [\ion{N}{II}],
[\ion{S}{II}], [\ion{Ni}{II}] and [\ion{Cr}{II}] have also been
identified.  The Balmer lines H$\alpha$, H$\beta$ and H$\gamma$
show P-Cygni behaviour clearly indicating post-AGB mass-loss
process in the object with the wind velocity up to 170 \kms.

\end{abstract}


\section{INTRODUCTION}\label{sec:intro}

Low- and intermediate-mass stars ($M_{\text{ZAMS}} \sim
0.8-–8M_{\odot}$) evolving from the asymptotic giant branch (AGB)
to become planetary nebulae (PNe) pass through a short lived but
important evolutionary stage that is designated as post-asymptotic
giant branch (post-AGB). Among post-AGB stars, there is a small
group of hot objects, early B supergiants with emission lines in
the spectrum, that are presumed to be the immediate progenitors of
the central stars of planetary nebulae (CSPN). The temperatures of
these stars are already high enough  for the ionization of their
surrounding envelopes to begin, but the ultraviolet radiation is
still insufficient for the excitation of [\ion{O}{III}] lines
typical of PNe.

The star LS 5112 from the Catalog of Luminous stars in the
Southern Milky Way by  \citet{ss71}  was identified with the
infrared (IR) source IRAS~18379--1707 from  catalogue of new
possible planetary nebulae \citep{PM1988} and was classified as a
hot post-AGB star by \citet{pvd00} with a spectral type of B1IIIpe
and IRAS colours typical of PNe. The star is contained in The
Toru\'{n} catalogue of Galactic post-AGB and related objects of
\citet{Szczerba07}. \citet{Nyman92} did not find circumstellar
CO(1-2) for IRAS 18379--1707. The object is not detected in the
radio at 3.6 cm \citep{umana04}. Combining the optical, near and
far-IR (\textit{ISO}, \textit{IRAS}) data of IRAS~18379--1707
\citet{gp04} have reconstructed its spectral energy distribution
(SED) and estimated the star and dust temperatures, mass loss
rates, angular radii of the inner boundary of the dust envelopes
and the distances to the star. For IRAS~18379--1707 \citet{c09}
detected a dual-chemistry circumstellar envelope, associated with
the 10 {\mkm} feature and silicate features due to PAHs. These
authors also present a DUSTY model of the continuum and SED and
derived a stellar temperature. IRAS 18379--1707 is
\ion{H}{}$_2$-emitting object. For the first time \ion{H}{}$_2$
emission from IRAS 18379--1707 is detected by \citet{hks04}.
\citet{kh05} assumed that \ion{H}{}$_2$ is excited by a mixture of
radiative and collisional excitation. The detailed structure of
the \ion{H}{}$_2$ nebula was shown by \citet{gf15}. The
\ion{H}{}$_2$ nebula takes the form of an oval shell of dimensions
3.6$\times$2.2 arcsec, whereas \ion{Br}{}$\gamma$ and \ion{He}{I}
emission is centrally located and spatially unresolved, indicating
a still-compact ionized region with densities $n_H\sim10^5$
cm$^{3}$ \citep{gf15}.

Some stellar and dust parameters of IRAS~18379--1707 are
summarized in Table~\ref{param}.

So far only low-resolution optical spectroscopy was performed for
IRAS~18379--1707. In this paper we report an analysis of the
high-resolution spectrum, on the basis of which the chemical
composition was obtained for the first time and the fundamental
parameters of the star with the best accuracy at the moment were
determined. Because of the high-resolution spectrum we could
resolve, identify and analyse many absorption lines, emission
lines, and P-Cygni profiles, etc. Because of the high spectral
resolution we could measure the radial velocities accurately and
discovered that LS~5112 is a high velocity star.

The paper is organized as follows: in Sect.~\ref{sec2} we describe
the observations and the data reduction; in Sect.~\ref{sec3}  we
present an analysis of the main spectral features; the estimation
of atmospheric parameters and abundances are presented in
Sect.~\ref{sec4}. In Sect.~\ref{sec5} we analyze the emission
spectrum and discuss our results in the context of post-AGB evolution
in Sect.~\ref{sec6}. In Sect.~\ref{sec7}, we give conclusions.

\begin{table*}
 \begin{center}
  \caption{Stellar and Dust Parameters of IRAS~18379--1707.}
 \label{param}

 \begin{tabular}{ccc}
 \hline
  Quantity                 &Value                                        & References\\
 \hline
 Position (J2000.0)        &$\alpha$=18:40:48.62 $\delta$=--17:04:38.33  & SIMBAD\\
 Gal. coord.               &$l$=016.50 $b$=--05.42                       & SIMBAD\\
 Parallaxes (mas)          &0.2593$\pm$0.0648                            & G18\\
 Distance (kpc)            &3497$_{-703}^{+1120}$                        & BJ18\\
 Spectral type             &OB+e, B2.5Ia, B1IIIpe                        & SS71, V98, P00\\
 $T_{\text{eff}}$ (K)      &19~000, 24~000                               & GP04, C09\\
 $\log g$                  &$2.5\pm0.5$                                  & GP03\\
 Magnitude                 &14.980(FUV) 14.599(NUV)                      & B-AGdC16\\
                           &11.94($U$) 12.38($B$) 11.93($V$)             & R98\\
                           &10.76($J$) 10.55($H$) 10.33($K$)             & GL97\\
                           &10.661($J$) 10.433($H$) 10.155($K_s$)        & 2MASS\\
 $E(B-V)$ (mag)            &0.71                                         & GP04\\
 $T_{\text{dust}}$ (K)     &140                                          & GP04\\
                           &120, 590                                     & C09\\

  \hline
 \end{tabular}
 \end{center}
 References. SIMBAD: \url{http://simbad.u-strasbg.fr/simbad/}; G18: \citet{gaia18}; BJ:
 \citet{bj18}; SS71: \citet{ss71}; V98: \citet{v98}; P00: \citet{pvd00};
 GP04: \citet{gp04}; C09: \citet{c09}; GP03: \citet{gp03}; B-AGdC16: \citet{bagdc16}; R98: \citet{r98};
 GL97: \citet{gl97a}
 \end{table*}

\section{OBSERVATIONS AND DATA REDUCTION}\label{sec2}

Two high-resolution optical spectra of IRAS~18379--1707 were
acquired on April 17, 2006 with the Fiber-fed Extended Range
Optical Spectrograph (FEROS) \citep{kauf99}, attached to the
MPG/ESO 2.2-m telescope at La Silla Observatory, Chile (Prop.ID:
77.D-0478A, PI: M. Parthasarathy). FEROS is a bench-mounted
echelle spectrograph, which provides data with a resolving power
$R\sim 48\,000$ and a spectral coverage from from 3600 to 9200
\AA\ in 39 orders. An EEV 2k$\times$4k CCD detector with a pixel
size of 15 {\mkm} was used. The exposure time of each spectrum was
2700 s. The a signal-to-noise (S/N) ratio was 100 per pixel in the
5500 \AA\ region. The reduction process was performed using the
FEROS standard on-line reduction pipeline and the echelle spectra
reduction package ECHELLE in IRAF using a standard reduction
manner including bias subtraction, removing scattered light,
detector sensitivity correction, removing cosmic-ray hits, airmass
extinction correction, flux-density calibration, and an all
echelle order connection. Both reduced spectra were continuum
normalised, co-added and cleaned of telluric lines with {\small
MOLECFIT} \citep{Kausch}.

\section{DESCRIPTION OF THE HIGH RESOLUTION SPECTRUM}\label{sec3}

The optical spectrum of IRAS~18379--1707 displays stellar
absorption lines, nebular emission lines and interstellar
absorption features. The identification of the emission lines in
spectrum of IRAS~18379--1707 are based on the Moore multiplet
table \citep{moore45} and the National Institute of Standards and
Technology (NIST) Atomic Spectra
Database\footnote{\url{https://www.nist.gov/pml/atomic-spectra-database}}.

The complete continuum-normalised spectrum of IRAS~18379--1707 in
the spectral ranges 3700--9200 \AA\ is presented at
\url{http://lnfm1.sai.msu.ru/~davnv/ls5112}.

\subsection{Photospheric absorption lines}

Absorption lines of neutral species including \ion{H}{i},
\ion{He}{i}, \ion{C}{I}, \ion{N}{I}, \ion{O}{I}  and \ion{Ne}{I}
were identified. Singly-ionized species including \ion{C}{II},
\ion{N}{II}, \ion{O}{II}, \ion{Si}{II}, \ion{S}{II} and
\ion{Mg}{II} were detected. Higher ionization is seen in
\ion{Al}{III}, \ion{Si}{III}, \ion{S}{III},  and \ion{Si}{IV}.

\subsection{Nebular emission lines}\label{emlines1}

 The list of emission lines in IRAS~18379--1707 is given in
 Table~\ref{emlines}. It includes the measured and laboratory
 wavelength (in the air), the equivalent width ($EW$), the
 heliocentric radial velocity ($V_\text{r}$), the name of the
 element and the multiplet number to which the measured line
 belongs. The hydrogen and helium lines  are not included in the Table~\ref{emlines}
 because they  have the complex multicomponent profiles and  will be discussed
 separately.

 The permitted  emission lines, in addition to hydrogen and helium,
 belong  to the ions of \ion{Si}{II}, \ion{S}{II}, \ion{C}{II}, \ion{Mg}{II}
 and also to the nonionised atoms of \ion{O}{I}, \ion{N}{I} and \ion{Na}{I}.
 Two weak emission lines  of \ion{Fe}{III} $\lambda$5126 and $\lambda$5156 are also present in the spectrum of
 IRAS~18379--1707. In the red spectral region, the permitted \ion{O}{i} $\lambda$8446 triplet is the most
 remarkable emission feature. The strong emissivity of this line as
 a fluorescence effect to the practically exact coincidence between L$\beta$ ($\lambda$1025)
 and \ion{O}{i} line at $\lambda$1026 was explained by \citet{bowen}.

 The forbidden emission lines are from [\ion{Fe}{II}], [\ion{Ni}{II}], [\ion{N}{II}] $\lambda$6548, 6584,
 [\ion{S}{II}] $\lambda$6717, 6731, [\ion{Cr}{II}] as well as
 and [\ion{O}{I}] $\lambda$5577, 6300, 6363.  The presence of the [\ion{N} {II}] and
 [\ion{S} {II}] emissions indicates the onset of the ionization of
 the circumstellar envelope and it is evolving towards the early stage of young
 low excitation planetary nebula.

\subsection{The hydrogen lines}

Balmer lines from \ion{H}{}15 to \ion{H}{}9 consist of the
photospheric absorption component and blue-shifted wind
absorption. \ion{H}{$\epsilon$} are blended with the interstellar
line of \ion{Ca}{II} at 3968 \AA. Pashen lines on our spectrum are
presented by high members from P10 and more. The line profile of
each of them consist of the photospheric absorption component and
weak emission on the blue wing of the absorption line.

\begin{figure}
\includegraphics[width=\columnwidth]{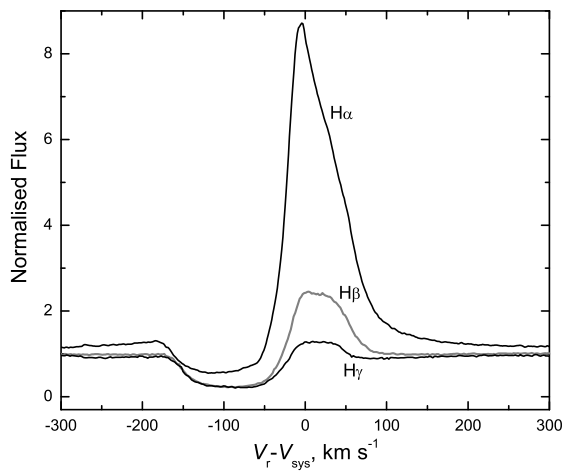}
\caption{P-Cygni profiles of H$\alpha$, H$\beta$ and H$\gamma$
lines seen in the spectrum of IRAS 18379--1707 on a velocity scale
relative to $V_{\rm sys} = -124$ \kms.} \label{balm}
\end{figure}

The profiles of the first Balmer lines H$\alpha$--H$\delta$
display a complex P-Cygni structure with its blue edge reaching a
value of up to --170 \kms. Fig.~\ref{balm} shows the
high-resolution spectra of the H$\alpha$, H$\beta$ and H$\gamma$.
The equivalent widths of the H$\alpha$ and H$\beta$ emission
component are 8.84 \AA\ and 1.61 \AA, respectively.

\subsection{The helium lines}

The \ion{He}{I} emission lines in IRAS~18379--1707 are superposed
on the corresponding absorption components. Fig.~\ref{hei} shows
the profiles of selected \ion{He}{i} lines and compared to model
spectra (see below). The asymmetric nature of the emission lines
suggests that they may have P-Cygni profiles.

\begin{figure*}
\includegraphics[scale=1.4]{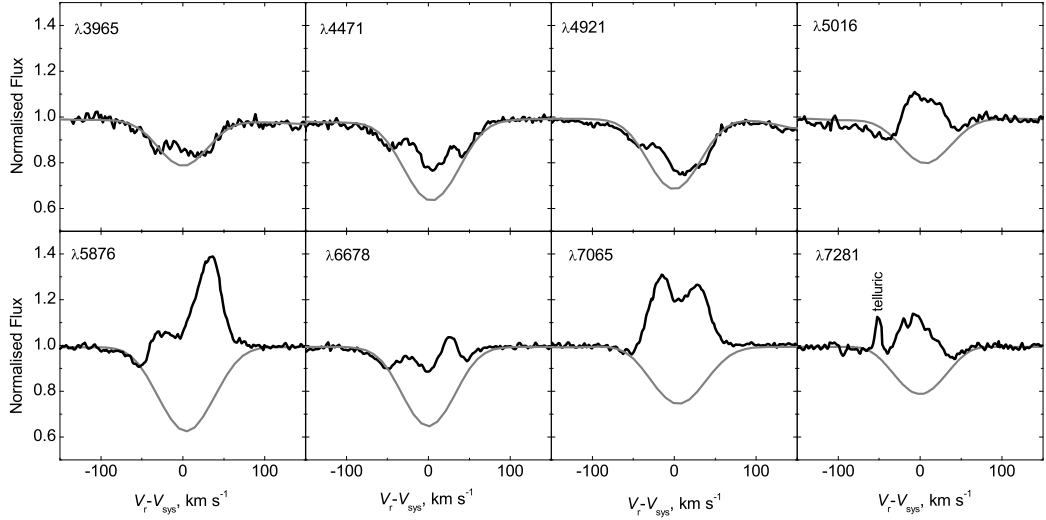}
\caption{The profiles of selected \ion{He}{i} lines seen in the
spectrum of IRAS 18379--1707 on a velocity scale relative to
$V_{\rm sys} = -124$ \kms and compared to model spectra (grey
lines).} \label{hei}
\end{figure*}

\subsection{Interstellar features and colour excess}

\begin{figure}
 \includegraphics[width=\columnwidth]{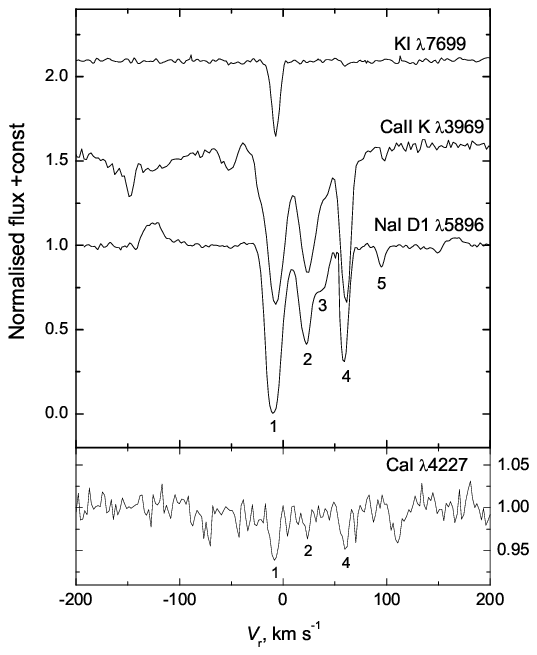}
 \caption{The profiles of the interstellar lines in
  the spectrum of IRAS~18379--1707. The various absorption
  components have been labelled.}
 \label{fig:IS}
\end{figure}

The spectrum of IRAS~18379--1707 contains absorption features that
have interstellar origin. There are \ion{Na}{I} doublet
($\lambda$5889.951, 5895.924), \ion{Ca}{II} H and K lines
($\lambda$3968.469, 3933.663), \ion{K}{I} lines
($\lambda$7664.899, 7698.974), and \ion{Ca}{I} at 4226.73 \AA. The
\ion{Na}{I}, \ion{Ca}{II} and \ion{Ca}{I} lines have
multi-component profiles, whereas the \ion{K}{I} lines show a
single and sharp profiles. H line of \ion{Ca}{II} is very much
blended with the strong stellar \ion{H}{$\epsilon$} feature. The
selected interstellar spectral lines are depicted in
Fig.~\ref{fig:IS}. Heliocentric radial velocities ($V_\text{r}$)
for the defined absorption components of \ion{Na}{I} D1 and D2,
\ion{Ca}{II} H and K lines are presented in Table~\ref{ISlines}.
The radial velocity of the \ion{K}{I} lines is close to the radial
velocity of {\lq}1{\rq} component of the \ion{Na}{I}, \ion{Ca}{II}
and \ion{Ca}{I} lines and is equal to --7.9$\pm$0.5 \kms. If one
compares the radial velocities of these components with the
average radial velocities for the star --124 \kms
(Sec.~\ref{subsec:Vr}), we may infer that all components in the
velocity interval from --10 to 100 \kms~originate in the
interstellar medium. \citet{Smoker04} found in the \ion{Ca}{II} K
and \ion{Ca}{II} H spectra of IRAS~18379--1707 an absorption
feature at $V_\text{LSR}\sim-137$ \kms. This component is also
present in our spectrum and it is most likely of interstellar
origin.

\begin{table*}
\begin{center}
 \caption{Absorption components of \ion{Na}{I} D1 and D2, \ion{Ca}{II} H and K
 lines in the spectrum of IRAS 18379-1707.
 $V_\text{r}$ are the respective heliocentric radial velocities.}
 {\small

\label{ISlines}
\begin{tabular}{ccccccccc}
  \hline
  component&\multicolumn{2}{c} {\ion{Na}{I} D$_{2}$} & \multicolumn{2}{c} {\ion{Na}{I} D$_{1}$}&
            \multicolumn{2}{c} {\ion{Ca}{II} H}      & \multicolumn{2}{c} {\ion{Ca}{II} K}\\

           & $\lambda_\text{obs.}$&$V_\text{r}$      & $\lambda_\text{obs.}$&$V_\text{r}$
           & $\lambda_\text{obs.}$&$V_\text{r}$      & $\lambda_\text{obs.}$&$V_\text{r}$ \\
           &     \AA              &     \kms         &    \AA               & \kms
           &     \AA              &     \kms         &    \AA               & \kms        \\
  \hline

  1&5889.75&--10.33&5895.73&--9.81&3968.36&--7.86&3933.57&--7.04\\
  2&5890.41&23.26&5896.41&24.76&3968.78&24.26&3933.98&24.19\\
  3&5890.71&38.53&5896.71&40.02&--&--&--&--\\
  4&5891.16&61.44&5897.13&61.37&3969.27&60.30&3934.45&60.16\\
  5&5891.91&99.61&5897.88&99.51&3969.75&96.73&3934.95&97.68\\

  \hline
\end{tabular}
}
\end{center}
\end{table*}

In addition to the above-mentioned interstellar atomic lines, our
echelle spectra contain several quite strong Diffuse Interstellar
Bands (DIBs) presented in Table~\ref{DIB}. It includes the
measured wavelength and the central wavelength from
\citet{Hobbs08}, the width (FWHM), the equivalent width ($EW$),
$EW/E(B-V)$ from  \citet{luna08}, $E(B-V)$ and the radial velocity
($V_\text{r}$). Three of DIBs centred at 6284, 6993, and 7224 \AA\
are strongly affected by telluric contamination. For these
features, the telluric component was removed before determining
their parameters.

As seen in Table~\ref{DIB} the radial velocities of most DIBs are
close to those of {\lq}1{\rq} or {\lq}2{\rq} components of the
\ion{Na}{I}, \ion{Ca}{II} interstellar lines. A radial velocity
analysis of the DIBs observed in IRAS~18379--1707 confirms our
result, as the Doppler shifts measured are found to be consistent
with an interstellar origin.

Using  $EW/E(B-V)$ from \citet{luna08} we estimated the extinction
by 8 DIBs and obtained the mean value $E(B-V)=0.61\pm0.08$ mag.
The DIB at 5849.81 \AA\ which is found to be unusually strong was
excluded from the consideration. The resulting $E(B-V)$ value is
close to the $E(B-V)=0.71$ mag obtained by \citet{gp03, gp04}.

An analysis of DIBs observed in IRAS~18379--1707 confirms the
conclusion of \citet{luna08} that, like in other post-AGB stars,
these features are of exclusively interstellar origin.

\begin{table*}
\begin{center}
 \caption{DIBs in the spectrum of IRAS~18379--1707.}
 {\small

\label{DIB}
\begin{tabular}{ccccccc}
  \hline

$\lambda_\text{obs.}$   & $\lambda_\text{c}$ &FWHM &$EW$ &$\frac{EW}{E(B-V)}$&$E(B-V)$&$V_{\text{r}}$\\
    \AA                 &       \AA             &\AA  &\AA  &\AA/mag            &mag     &\kms          \\

  \hline

4963.72 & 4963.88 & 0.62 & 0.02& --   & --  & -9.7\\
5488.10 & 5487.69 & 2.98 & 0.11& --   & --  & 22.4\\
5493.75 & 5494.10 & 1.44 & 0.04& --   & --  & 19.1\\
5705.02 & 5705.08 & 0.47 & 0.10& --   & --  & -3.2\\
5780.40 & 5780.48 & 1.92 & 0.33& 0.46 & 0.71& -4.2\\
5796.90 & 5797.06 & 0.95 & 0.10& 0.17 & 0.61& -8.3\\
5809.31 & 5809.23 & 0.72 & 0.03& --   & --  &  4.1\\
5849.68 & 5849.81 & 0.99 & 0.08& 0.061&1.30 & -6.7\\
6089.58 & 6089.85 & 0.80 & 0.02& --   & --  &-13.3\\
6195.78 & 6195.98 & 0.45 & 0.03& 0.53 &0.64 & -9.7\\
6203.11 & 6203.05 & 2.13 & 0.11& --   & --  &  2.9\\
6233.83 & 6234.01 & 0.66 & 0.04& --   & --  & -8.7\\
6269.73 & 6269.85 & 1.27 & 0.13& --   & --  & -5.7\\
6283.88 & 6283.84 & 2.85 & 0.45& 0.9  &0.50 & -3.3\\
6375.91 & 6376.08 & 0.55 & 0.05& --   & --  & -8.0\\
6379.08 & 6379.32 & 0.63 & 0.06& 0.088&0.74 &-11.3\\
6445.13 & 6445.28 & 0.49 & 0.03& --   & --  & -7.0\\
6613.44 & 6613.62 & 1.00 & 0.13& 0.21 &0.60 & -8.2\\
6660.44 & 6660.71 & 0.54 & 0.04& --   & --  &-12.2\\
6699.16 & 6699.32 & 0.54 & 0.02& --   & --  & -7.2\\
6992.90 & 6993.13 & 0.79 & 0.07& 0.12 &0.58 & -9.9\\
7116.16 & 7116.31 & 0.74 & 0.03& --   & --  & -6.3\\
7119.15 & 7119.71 & 0.86 & 0.08& --   & --  &-23.6\\
7223.78 & 7224.03 & 1.09 & 0.13& 0.25 &0.53 &-10.3\\

 \hline
\end{tabular}
}
\end{center}
\end{table*}

\section{DETERMINATION OF THE ATMOSPHERIC PARAMETERS}\label{sec4}

The stellar parameters ($T_{\rm eff}$, $\log g$, $v\sin i$,
$\xi_{\rm t}$, elemental abundances) are determined by fitting
synthetic line profiles to the observed ones. The synthetic
profiles were calculated with {\sc synspec}, using the BSTAR2006
grids generated with the code {\sc tlusty} \citep{hubeny95}, which
assumes a plane-parallel atmosphere in radiative, statistical
(non-LTE) and hydrostatic equilibrium. We used grids with scaled
solar abundances for metals $Z/Z_{\odot}=$0.5 and 0.2 and
microturbulent velocity of 10\kms, which are closest to the
obtained model parameters of IRAS~18379--1707. Differences in the
results, obtained on various grids, are analysed in
Sect.\,\ref{sec:err}.

Many lines show distortions in their profiles, as a rule it is an
emission feature in the blue wing (\ion{He}{i} lines, as example)
or appearance of an extended absorption blue wing (\ion{Ne}{i},
\ion{Si}{ii}, \ion{S}{ii}, \ion{Mg}{ii}). To obtain atmospheric
parameters, we used two approaches:
\begin{itemize}
\item Observed profile was fitted with synthetic one by $\chi^2$ minimization,
bad or distorted parts of the profile were ignored. Uncertainties of the
parameters for individual line were estimated from the obtained residual.
\item We compare $EW$ for synthetic and observed profiles.
\end{itemize}
For lines without visible distortions both approaches lead to the
same parameters within their uncertainties. Thus, the second
method makes sense only for lines with extended blue wing.
Finally, abundances derived from individual lines are averaged
with weights of their uncertainties.

The defined parameters are interconnected, therefore we obtain a
self-consistent set of parameters by iterations.

To test our methods and to check adequacy of the {\sc tlusty}
models for our object, we performed similar measurements for
ordinary blue supergiant 62 Ori with similar parameters.
Comparison with this star allows us to filter out artifacts
associated with an inaccuracy of our modelling of blue supergiants
in general from the features, specific for IRAS~18379--1707.

\subsection{Radial velocity}\label{subsec:Vr}

We selected 51 absorption lines (see Table\,\ref{tab:abslines}),
shapes of which are well fitted by theoretical ones. We did not
include lines with obvious distortions: emission features in
strong \ion{He}{i} lines; lines of relatively {\lq}cold{\rq} ions
\ion{Ne}{I}, \ion{Mg}{II}, \ion{S}{II}, which are blue-shifted by
$10-20$\,{\kms} relative to the most of the lines.
The wavelength shifts were found by fitting the line profiles with Gaussian for
both observed and calculated spectrum over the same wavelength
range. Measurements averaged over each ion are presented in
Table\,\ref{tab:radvel}. Weighted mean over all lines produces the
heliocentric velocity of the star $V_{\rm
r}=-124.0\pm0.4$\,{\kms}. Thus we conclude that IRAS~18379--1707
is a high velocity star. As we note lines with low excitation
potentials $E_l\lesssim17$ eV are blue-shifted (see
Table\,\ref{tab:blueshifted}), but we have not found any
dependence on excitation potential for the selected lines
($E_l\gtrsim17$ eV), nevertheless weighted scattering around mean
is $\sigma=2.8$\,{\kms} at average uncertainties of each
measurement of $\sim1$\,{\kms}.

The similar procedure was made for 62 Ori. We found that hydrogen
and helium lines of 62 Ori also show a blue shift $\sim10$\kms
relative to the other lines. We cannot exclude a such shift in
IRAS~18379--1707, but it is less pronounced in the line wings,
while the cores of the lines are distorted by strong P-Cygni
feature. The profiles of {\lq}cold{\rq} ions in 62 Ori do not show
peculiarities as in IRAS~18379--1707, but some of them are also
blue-shifted as a whole by $\sim10$\kms (see \ref{sec:chem} for
details).

\begin{table}
\caption{Radial velocities.}
 \label{tab:radvel}
 \begin{center}
\begin{tabular}{cccccc}
\hline
Ion          & $\overline{E}(\sigma_E)$& $\overline{V}+124$&  $\sigma_{\overline V}$& $\sigma_V$& N  \\
               &  eV                   &       {\kms}         &                   {\kms} &    {\kms}       \\
\hline
\ion{He}{i}    &  21(0)                &        -0.7       &      1.3               &      2.5  & 5  \\
\ion{C}{ii}    &  16(0)                &        -3.1       &      2.4               &      2.6  & 2  \\
\ion{N}{ii}    &  19(1)                &        -0.6       &      1.1               &      2.5  & 6  \\
\ion{O}{ii}    &  25(2)                &        -0.6       &      0.5               &      2.7  & 29 \\
\ion{Si}{iii}  &  21(3)                &         2.0       &      1.1               &      2.5  & 6  \\
\ion{Al}{iii}  &  16(0)                &        -1.2       &      3.6               &      3.9  & 2  \\
\ion{S}{iii}   &  18(0)                &        -0.0       &       --               &      1.3  & 1  \\
\hline
All            &                       &         0.0       &      0.4               &      2.8  & 51 \\
\hline
\multicolumn{6}{c}{Ions with blue-shifted lines} \\
\hline
\ion{Ne}{i}    &  17(0)                &       -10.0        &       4.3             &      7.5  & 4  \\
\ion{Mg}{ii}   &   9(0)                &       -13.4       &        --              &      1.0  & 1  \\
\ion{Si}{ii}   &  10(0)                &       -19.5       &       2.2              &      2.4  & 2  \\
\ion{S}{ii}    &  14(0)                &       -8.2        &       4.1              &      7.1  & 4  \\
\hline
\end{tabular}\\
\end{center}
\end{table}

\subsection{Surface gravity $\log g$ and effective temperature $T_\text{eff}$}\label{sec:tlgg}

\begin{figure}
\begin{center}
\includegraphics[scale=0.65]{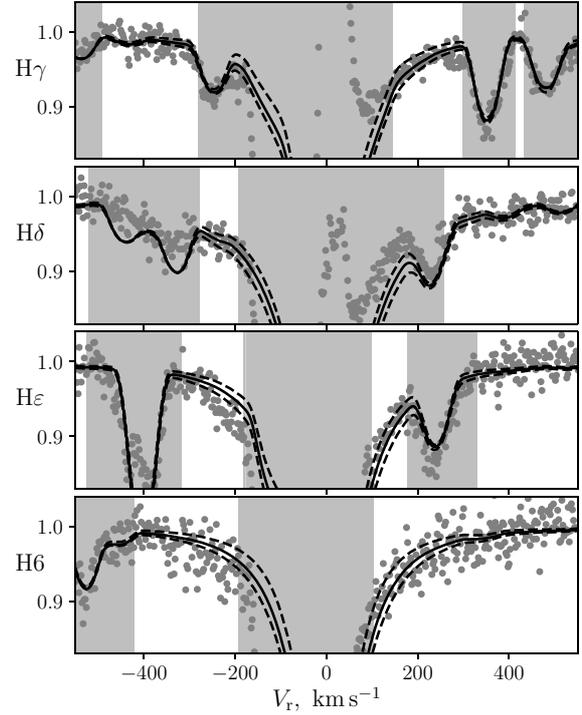}
\end{center}
\caption{Wings of hydrogen lines. The shaded areas are ignored
during the fit. Solid lines are for the best-fit values of $\log
g.$ Dashed lines indicate uncertainties in the profiles,
corresponding to 0.05 dex.}\label{hydprof}
\end{figure}

\begin{figure}
\begin{center}
\includegraphics[scale=0.65]{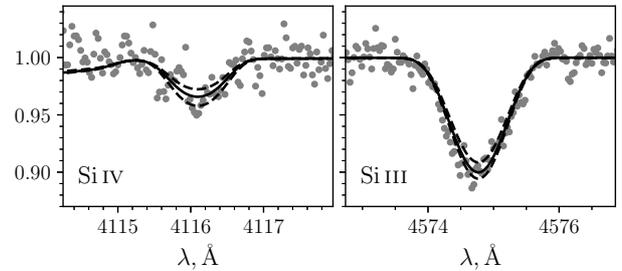}
\end{center}
\caption{Examples of silicon lines.
The solid line is for $T_{\rm eff}=18000$\,K,
dashed lines correspond to the uncertainty of $\pm300$\,K.}\label{si_prof}
\end{figure}

\begin{figure}
\begin{center}
\includegraphics[scale=0.65]{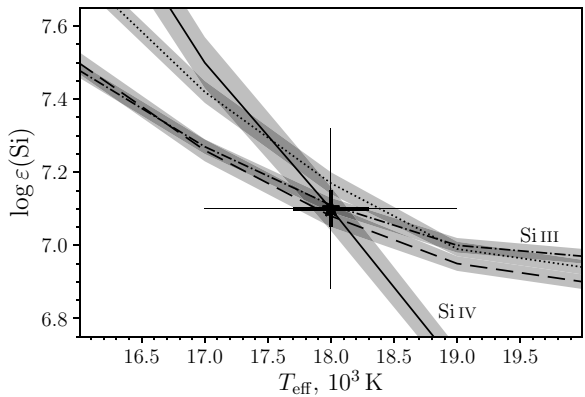}
\end{center}
\caption{The silicon ionization balance. The solid line is for
\ion{Si}{iv}\,$\lambda$4116.1; the dashed, dash-dotted and dotted
lines are for \ion{Si}{iii}\,$\lambda$4567.8, 4574.8, 5739.7. The
shadowed regions show estimated uncertainties for each line. The
star with errorbars marks the intersection point and its
uncertainty: the thick line is for uncertainties related with the
quality of the fit, the thin line corresponds to the total error,
which includes uncertainties in other variables and model
limitations, see Sect.\,\ref{sec:err}.}\label{tlgg}
\end{figure}

Surface gravity $\log g$ along with $T_{\rm eff}$ was determined
from wings of hydrogen lines and from the silicon ionization
balance (\ion{Si}{iii}/\ion{Si}{iv}). H\,$\alpha$ and H\,$\beta$
are strongly distorted by P-Cygni features even in far wings, so
we used H\,$\gamma,$ H\,$\delta,$ H\,$\varepsilon$ and H\,6, {the
stark broadening for which was accounted according to Lemke tables
\citep{lemke97}.} We found that the synthetic profiles of hydrogen
lines fit observations (see Fig.\,\ref{hydprof}) for any pairs of
$\log g$ and $T_{\rm eff},$ which satisfy the equation $\log
g=2.44\pm0.05+9\times10^{-5}(T_{\rm eff}(K)-20\,000).$ For these
pairs of $\log g$ and $T_{\rm eff},$ we fit silicon lines,
adjusting the abundance $\varepsilon(\rm Si)$ for each line
individually. To do this, we selected only non-blended lines:
\ion{Si}{iii}\,$\lambda$4567.8, 4574.8, 5739.7, and
\ion{Si}{iv}\,$\lambda$4116.1 (see Fig.\,\ref{si_prof} for
examples). The obtained results are presented in Fig.\,\ref{tlgg},
from which we can see that the abundances measured from
\ion{Si}{iii} and \ion{Si}{iv} lines are in agreement with each
other at $T_{\rm eff}=18\,000\pm300$\,K and $\log g=2.25\pm0.05$
and equal to $\log \varepsilon(\rm Si)=7.10\pm0.05.$

The observed spectrum contains also lines of \ion{Si}{ii},
\ion{S}{ii} and \ion{S}{iii}, which could be used for
determination of stellar parameters from the ionization balance of
\ion{Si}{ii}/\ion{Si}{iii} and \ion{S}{ii}/\ion{S}{iii}, however
lines of {\lq}cold ions{\rq} like \ion{Si}{ii}, \ion{S}{ii} are
blue-shifted with respect to the most of the lines, i.e. they
originate in an outflowing gas, which is not accounted in the
hydrostatic {\sc tlusty} models. Up to the date the atmospheric
parameters for post-AGB stars were based on the hydrostatic
LTE/non-LTE models, because there are not adequate models
applicable for determination of parameters for the outflowing
atmospheres for these stars. Although the obtained parameters have
not a strict sense, if we use lines without distortions, the
deviations from the true values are suspected to be small (see
discussion in \citealt{mello12}).

The same procedure was applied for 62 Ori. In this case we obtain
$T_{\rm eff}=17600$\,K, $\log g = 2.15$, $\log \varepsilon(\rm
Si)=7.52$. Our value of $T_{\rm eff}$ is less than previous
estimates $T_{\rm eff}=19000\pm1000$\,K obtained by
\citet{haucke18} with the code {\sc fastwind} and by
\citet{crowther06} with {\sc cmfgen}. Unlike IRAS~18379--1707,
lines of \ion{Si}{ii} do not show deviations and result in the
same $T_{\rm eff}$ and $\varepsilon(\rm Si)$, which follow from
\ion{Si}{iii/iv} balance.

\subsection{Microturbulence $\xi_{\rm t}$}\label{sec:vturb}

\begin{figure}
\begin{center}
\includegraphics[scale=0.65]{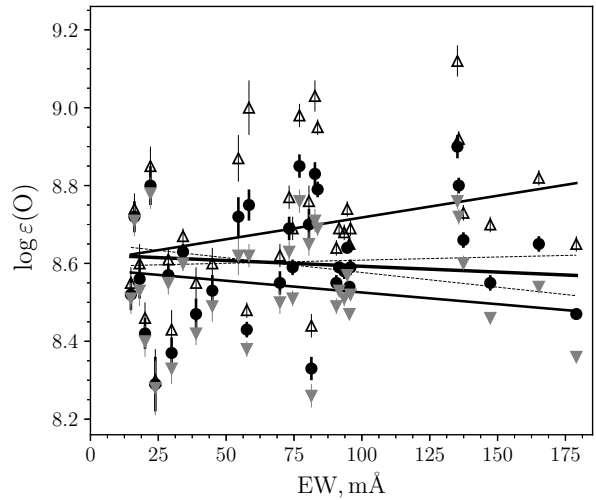}
\end{center}
\caption{ The abundance--equivalent width diagram to determine the
microturbulence velocity. Empty triangles, filled circles, and grey triangles
show measurements of $\log \varepsilon({\rm O})$ at  $\xi_{\rm t}=7,$
10, 13\,{\kms}, correspondingly.
The solid lines are linear fits found by the error-weighted least-squares,
the thick one corresponds to $\xi_{\rm t}=10$\,{\kms} and the dashed lines around it
illustrate the uncertainties in the slope coefficient.
}\label{vmic}
\end{figure}

The microturbulence velocity was derived from the analysis of 34
\ion{O}{ii} lines non-blended with lines of other ions. For each
line we adjust oxygen abundance $\varepsilon({\rm O})$ for three
trial values $\xi_{\rm t}=$7, 10, 13\,{\kms}. The obtained
dependencies of $\varepsilon({\rm O})$ on $EW$ of the lines are
shown in Fig.\,\ref{vmic}. Slopes of these dependencies were
calculated by the error-weighted least-squares. The
microturbulence of 7\,{\kms} produces the slope of $k=1.1\pm0.5,$
while $k=-0.3\pm0.5$ for $\xi_{\rm t}=10$\,{\kms} and
$k=-0.6\pm0.5$ for $\xi_{\rm t}=13$\,{\kms}. Interpolating between
these values, we obtain that zero slope should be achieved at
$\xi_{\rm t}=9^{+3}_{-1}$\,{\kms}. For simplicity we accept
$\xi_{\rm t}=10\pm2$\,{\kms}. The weighted mean of $\log
\varepsilon({\rm O})=8.59$ with a standard error of 0.02\,dex and
a standard deviation of 0.11\,dex.

For 62 Ori we obtain $\xi_{\rm t}\sim30$\,{\kms} and $\log \varepsilon({\rm O})=8.68$
with a scatter of 0.04\,dex instead of 0.11 dex for IRAS~18379--1707.

\subsection{Chemical abundances}\label{sec:chem}

Elemental abundances were determined for the following model
parameters: $T_{\rm eff}=18\,000$\,K, $\log g=2.25,$ $\xi_{\rm
t}=10${\kms}. The obtained results are collected in
Table\,\ref{tab:abund} and commented below. Uncertainties in the
abundances are mostly related to the real irreducible differences
between the synthetic and observed spectra rather than the quality
of the observations. We suppose that these differences arise due
to inaccurate atomic data as well as a noticeable difference
between the real, possibly non-hydrostatic, stellar atmosphere and
the {\sc tlusty} models. However, we emphasize that spectrum of
the blue supergiant is better reproduced by {\sc tlusty} models
than the spectrum of IRAS~18379--1707. Both objects have similar
parameters and show signatures of an outflowing atmosphere,
therefore limitations of our modelling should have appeared in
both cases. Probably the key difference is a higher mass loss for
IRAS~18379--1707, which is significantly more luminous in
H$\alpha$.

\begin{table}
\caption{Element abundances determined for IRAS~18379--1707 in
$\log \varepsilon=12+\log(n_\text{X}/n_\text{H})$ and
[X/H]=$\log(n_\text{X}/n_\text{H})-\log(n_{\text{X}\odot}/n_{\text{H}\odot})$}
 \label{tab:abund}
 \begin{center}
\begin{tabular}{lcccccc}
\hline
        & $\log \varepsilon_{\odot}$   & $\log \varepsilon$ & [X/H]& $\sigma_x$&$\sigma_{\overline{x}}$ &N  \\
\hline
C       &   8.43  &   7.76           &--0.67   &  0.03     &   0.03      & 2  \\
N       &   7.83  &   7.41           &--0.42   &   0.09    &   0.04      & 7 \\
O       &   8.69  &   8.59           &--0.10   &   0.11    &   0.02      & 34 \\
Ne      &   7.84  &   8.39*          &  0.55   &   0.17   &   0.06       & 8 \\
Mg      &   7.60  &   7.30*          &--0.30   &   --     &   0.15       & 1 \\
Al      &   6.45  &   5.56           &--0.89   &   0.02    &   0.02      & 2 \\
Si      &   7.51  &   7.10           &--0.41   &   0.05    &   0.03      & 4 \\
S       &   7.12  &   6.39           &--0.73   &   0.05    &   0.05      & 2 \\
Fe      &   7.50  &   $\lesssim7.5$  &$<0$     &    --     &    --       & --\\
\hline
\end{tabular}
\end{center}
Solar values $\log \varepsilon_{\odot}$ are taken from
\citet{asplund09}. $\sigma_x$ and $\sigma_{\overline{x}}$ are the
standard deviation and the standard error of mean for $\log
\varepsilon.$ N is a number of averaged lines. Abundances marked
with * were measured on lines with extended blue wing and probably
biased.
\end{table}

\subsubsection*{He}
Almost all helium lines have distortions of the profiles. We
selected 10 lines, in which these distortions are minor, and
derived the helium abundance by fitting wings of these lines.
Regardless of method of abundance determination (from profile
fitting or from $EW$) the helium lines show dependence of
$\varepsilon$ on $EW$, as if they are formed in gas with $\xi_{\rm
t}({\rm He})=30$\,\kms that is significantly higher than $\xi_{\rm
t}({\rm O})=10$\,\kms, derived from oxygen lines, moreover helium
turns to be underabundant $\log\varepsilon({\rm O})=10.65.$ We
suppose that these parameters are unrealistic, and were obtained
due to various distortions in profiles, in particular, due to
unaccounted emission, filled-in photospheric lines, such as
directly observed in strong lines (see Fig.\, \ref{hei}).

\subsubsection*{C}
In the case of 62 Ori, carbon abundances derived from \ion{C}{ii}
$\lambda$3921, 4227 and 6578 lines are consistent with each other
and result in $\log \varepsilon(\rm C) = 7.60.$ The line at 3919
\AA\ is blended with nitrogen, which is enhanced for this star,
therefore we exclude this line. Both lines \ion{C}{ii}
$\lambda$6578/6583 are blue-shifted, and the line at 6583 \AA\
gives $\log\varepsilon({\rm C})=7.70.$ \citet{crowther06} using
{\sc cmfgen} models have found $\log\varepsilon({\rm C})=7.65.$
Therefore, {\sc tlusty} model can reproduce the carbon spectrum of
62 Ori  and results are consistent with previous study.

\begin{figure}
\begin{center}
\includegraphics[scale=0.65]{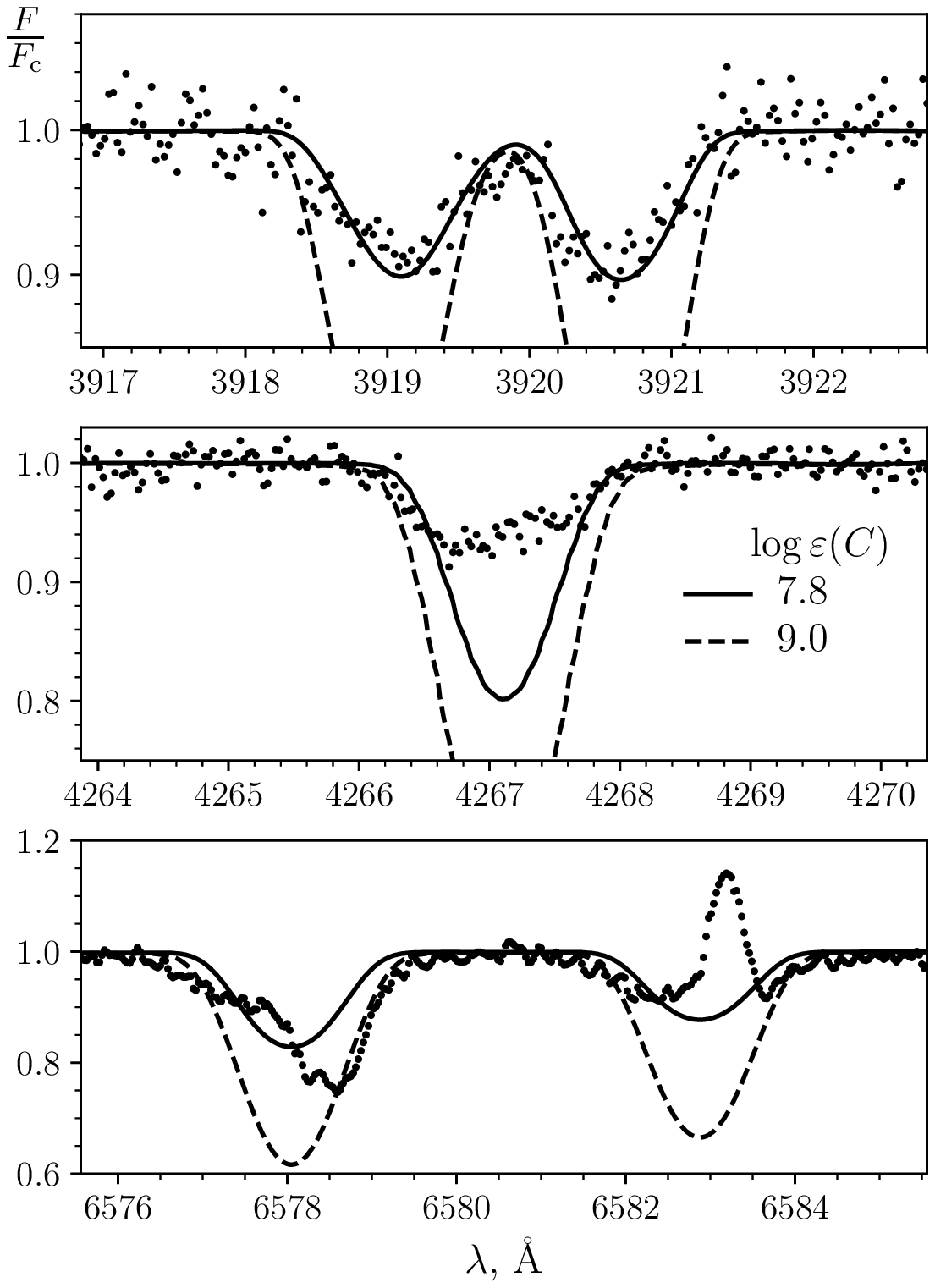}
\end{center}
\caption{\ion{C}{ii} line profiles. Dots are for the observations,
lines are for synthetic profiles for various $\log\varepsilon(C).$}\label{fig:carbon}
\end{figure}

In the case of IRAS~18379--1707, all mentioned lines show large
distortion (see Fig.\,\ref{fig:carbon}), except doublet $\lambda$3919, 3921,
which gives $\log\varepsilon({\rm C})=7.76.$ The line at 4227 \AA\ is also consistent
with this value, bearing in mind signatures of emission core in its centre.

The line at 6578 \AA\ has a complex shape, which can be interpreted
either as a shallow stellar absorption with an additional more narrow
absorption feature in the red wing, or as a deep stellar absorption
with an emission at the blue wing (see Fig.\,\ref{fig:carbon}).
In the last case we need very high carbon
abundance $\log\varepsilon({\rm C})\sim 9,$ which is in contradiction with other lines.
The profile of \ion{C}{ii} $\lambda$6583 is blended with the
forbidden line \ion{N}{ii} $\lambda$6584. A joint analysis of
the profiles of \ion{C}{ii} $\lambda$6578/6583 and [\ion{N}{ii}]
$\lambda$6548/6584 shows that the observed narrow absorption
feature in \ion{C}{ii} $\lambda$6578 must also be present in
\ion{C}{ii} $\lambda$6583, but it is almost completely covered by
[\ion{N}{ii}] $\lambda$6584. Therefore, both components $\lambda$6578/6583 show
profiles with the same distortions.

\subsubsection*{N}
The nitrogen abundance is measured from 7 individual \ion{N}{ii}
lines. There are not any significant dependence between $\log
\varepsilon(\rm N)$ and $EW$ that means that the \ion{N}{ii} lines
consistent with the accepted value $\xi_{\rm t}.$ We do not use 4
more lines, which give unreasonable results for 62 Ori. Nitrogen
abundance for 62 Ori $\log\varepsilon=8.55$ coincides with
previous study by \citet{crowther06}:  $\log\varepsilon=8.55.$

\subsubsection*{O}
The oxygen abundance was determined along with the microturbulence
using 34 \ion{O}{ii} lines, see Sect.\,\ref{sec:vturb} and
Fig.\,\ref{vmic} for details. The obtained value of $\log
\varepsilon(\rm O)$ is $8.59\pm0.02.$ Both stars IRAS~18379--1707
and 62 Ori show similar scatter of $\varepsilon$ measured from
various lines $\sigma_{\varepsilon}\approx0.1$ dex. The mean value
of $\varepsilon$ for 62 Ori coincides with the solar 8.69, but
higher than value 8.45 deduced by \citet{crowther06}.

It should be noted that the \ion{O}{I} triplet lines at 7771-5
\AA\ are very strong ($EW=2.18$ \AA) indicating an extended
atmosphere and non-LTE effects. \ion{O}{I} $\lambda$6156, 6157
are enhanced in observations and show extended blue wing.

\subsubsection*{Ne}
We selected 8 lines of \ion{Ne}{i}. Strong lines show an extended
blue wing, such distortions are possible in weak lines, however
they are undetectable due to noise. Fitting central and red parts
of the profiles gives an abundance of $8.27\pm0.05,$ the abundance
derived from $EW$ measured over the whole profile is
$8.40\pm0.06.$ Neon lines in 62 Ori are significantly weaker and
blue-shifted by $\sim10$\,\kms, but without enhanced blue wing.
Abundance is higher than the solar one by only 0.1 dex and equals
to $7.96\pm0.03.$ Thus, enhancement of neon lines and distortion
in their profiles are observed only in IRAS~18379--1707, but
absent in the blue supergiant 62~Ori.

\subsubsection*{Mg, Al}

There are two \ion{Al}{iii} lines at 5697, 5723 {\AA}, which give
$\log\varepsilon=5.58\pm0.02$ and $5.54\pm0.02.$
One another line \ion{Al}{iii} $\lambda$4480 is weak and blended
with strong \ion{Mg}{ii} line at 4481 \AA, however it probably
requires a higher abundance by $0.15-0.2$\,dex.

The similar picture is seen in 62 Ori, where \ion{Al}{iii}
$\lambda$5697, 5723 give $\log\varepsilon=6.29$ and $\log\varepsilon=6.32,$
while \ion{Al}{iii} $\lambda$4481 better describes with $\log\varepsilon\approx6.5,$
which is consistent with solar value 6.45. We note that profiles
of \ion{Al}{iii} in 62 Ori are slightly distorted: the blue
wing is steeper than the red one.

We can conclude that Al abundance in 62 Ori is consistent
with the solar value, but in IRAS~18379--1707
it is certainly underabundant at least by 0.7 dex.

The \ion{Mg}{ii} $\lambda$4481 shows an extended blue wing, if we
fit the red and central part of the profile, we obtain $\log
\varepsilon(\rm Mg)=7.0\pm0.02,$ and $\log \varepsilon(\rm
Mg)=7.3,$ if we measure it from EW. The presence of a strong
distortion of the profile does not allow us to say that this
abundance is correct. In the case of 62 Ori \ion{Mg}{ii}
$\lambda$4481 show blue-shifted profile by approximately
$7$\,\kms, abundance derived from this line $\log \varepsilon(\rm
Mg)=7.3$ (the solar value is 7.6).

\subsubsection*{Si}
The silicon abundance was determined simultaneously with $T_{\rm
eff}$ based on three \ion{Si}{iii} lines and one \ion{Si}{iv}
line, see sect.\,\ref{sec:tlgg} and Fig.\,\ref{tlgg} for details.
The obtained value of $\log \varepsilon(\rm Si)$ is $7.1$ with
scatter of 0.05 and the standard error of mean 0.03.

\subsubsection*{S}

The observed spectrum contains lines of \ion{S}{ii} and
\ion{S}{iii}, but the \ion{S}{ii} $\lambda$5454, 5640, 5647 lines
show extended blue wings, at the same time there are the lines
(\ion{S}{ii} $\lambda$5212, 5322, 5346), which are predicted by
{\sc tlusty}, but absent in observations. We suppose that observed
\ion{S}{ii} lines are formed in the outflowing gas and cannot be
used for the abundance measurements without a proper model of the
expanding atmosphere. The sulfur abundance was measured from the
\ion{S}{iii} $\lambda$4254, 4285 lines, which show symmetric
profiles without signs of the outflow.

In the case of 62 Ori we use lines \ion{S}{iii} $\lambda$4254,
4285 and \ion{S}{ii} $\lambda$5454 and obtain $\log
\varepsilon(\rm S)=6.9\pm0.1$.

\subsubsection*{Fe}

The synthetic spectrum predicts lines of \ion{Fe}{iii} at solar abundance of Fe.
The absence of these lines in the observed spectrum can impose an upper limit
$\log \varepsilon(\rm Fe)<6.8.$
It should be noted that \ion{Fe}{iii} $\lambda$5127 and $\lambda$5156 lines appear
in the observed spectrum as emission lines.

In the case of 62 Ori \ion{Fe}{iii} $\lambda$4005, 4022 give
$\log \varepsilon({\rm Fe})=7.0$ and 7.1. More strong lines \ion{Fe}{iii}
$\lambda$5127, 5156 show peculiarities at the line centre
(it has more sharp shape than other lines) and lead
to $\log \varepsilon({\rm Fe})=7.8$ and 8.0.
If we assume that 62 Ori has near-solar Fe abundance and the lines $\lambda$4005, 4022
underestimate abundance for both stars nearly equally,
then we can conclude that IRAS~18379--1707 has $\log \varepsilon(\rm Fe)$
less than the solar value.

\begin{figure}
\begin{center}
\includegraphics[scale=0.65]{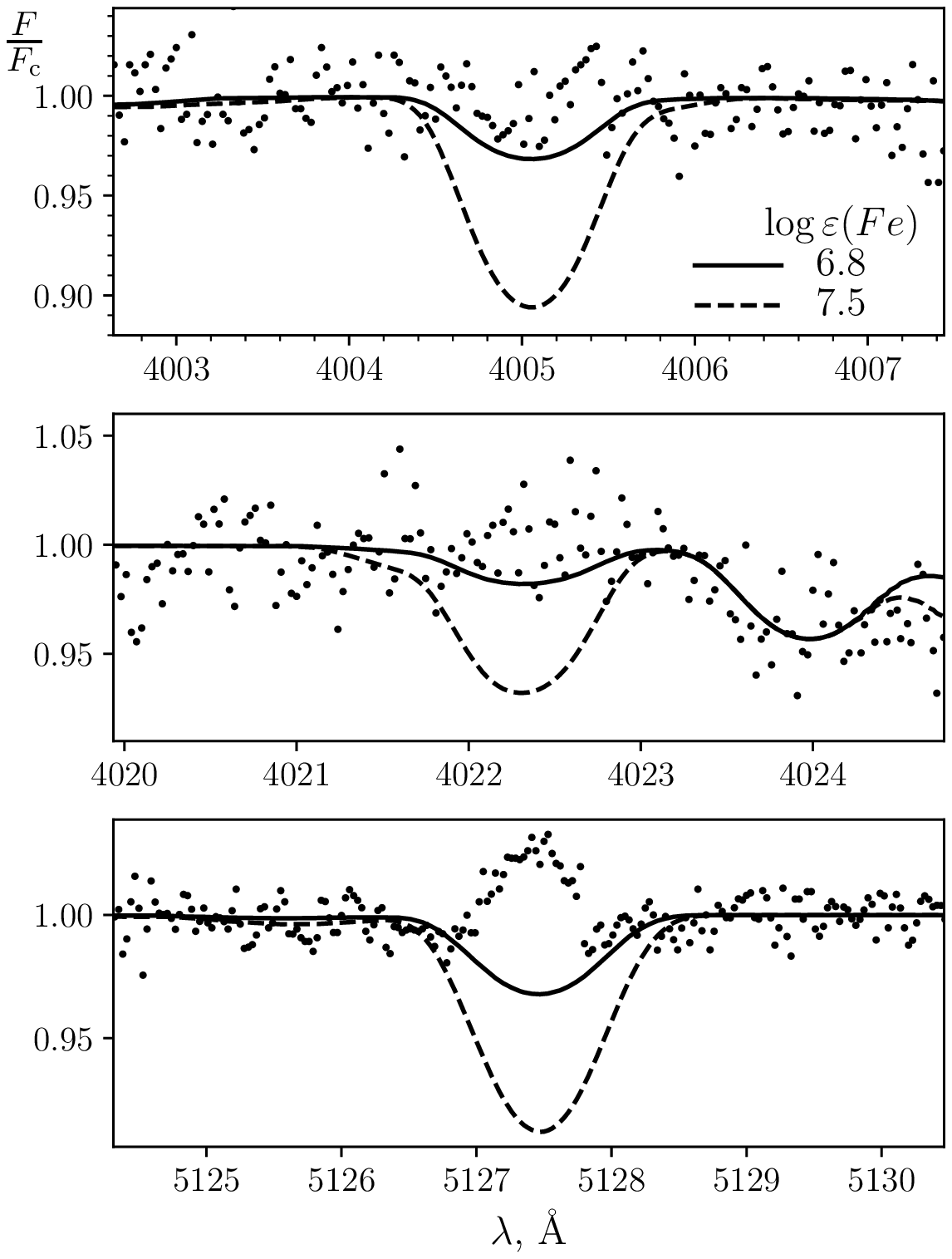}
\end{center}
\caption{\ion{Fe}{iii} line profiles. Dots are for the observations,
lines are for synthetic profiles for various $\log\varepsilon(\rm Fe).$
}\label{fig:ferrum}
\end{figure}

\subsection{Error analysis}\label{sec:err}

Uncertainties of the parameters estimated above reflect only
the goodness of fit of the observational data. To account
uncertainties and limitations in the modelling,
we estimate final uncertainties in the parameter $x_k$ as
$$\overline{\sigma}_k^2 = \sum\limits_{i}{\left( \frac{\partial x_k}{\partial x_i}\sigma_{i}\right)^2}
+ \sigma_C^2+ \sigma_M^2+\sigma_p^2+\sigma_{\xi m}^2,$$
where $\frac{\partial x_k}{\partial x_i}$ reflects change in $x_k$ when varying $x_i$
at fixed other parameters, these derivatives are collected in Table~\ref{tab:err}.
$\sigma_{i}$ is error in $x_i$ related with quality of the fit.
In addition to uncertainties of free parameters, we add uncertainties related with
continuum placement ($\sigma_{\text{C}}$); metallicity of the model grid ($\sigma_{\text{M}}$):
we explore how our results change between models with scaled abundances 0.2 and 0.5
of the solar values; inclusion or not turbulent pressure term in the hydrostatic equation
($\sigma_p$); differences in thermal structure of the atmosphere calculated for various
values of microturbulence within $\sigma_{\xi}=3$\,\kms ($\sigma_{\xi m}$).

\begin{table*}
\caption{The derived parameters, their sensitivity to the model
assumptions and final uncertainties.}
 \label{tab:err}
 \begin{center}
\begin{tabular}{l|rrr|rrrr|rrr}
\hline

                  &    $\partial x_i/ \partial T_{\rm eff}$    & $\partial x_i/ \partial \log g$ & $\partial x_i/ \partial \xi$  & $\sigma_{\rm C}$    & $\sigma_{\rm M}$ & $\sigma_{p}$ & $\sigma_{\xi m}$  & $x_i$ & $\sigma_{i}$ &$\overline{\sigma}_{k}$ \\
\hline
$T_{\text{eff}}$              &        1                                &  7200                        & -177                       & 540   &  400     &   160          &  270 & 18000 & 300            &  1000       \\
$\log g    $                  &     $10^{-4}$                           &    1                         &   0                        &  0     & 0.02    &   0.04         &  0.03 & 2.25  & 0.05           &  0.08       \\
$\xi       $                  &  $-3.3\times10^{-3}$                    &    0                         &   1                        &  0     &  2      &    1          &  1.8  & 10    &  2             &  3.6        \\
\hline
$\log \varepsilon(\text{C}) $ & $3.5\times10^{-5}$                      & -1.9                         & -0.016                     & 0.06  &   0.07  &    0.08        & 0.01 & 7.76 &   0.03         & 0.16      \\
$\log \varepsilon(\text{N}) $ & $-2.5\times10^{-5}$                     & -0.7                         & -0.003                     & 0.03  &   0.03  &    0.05        &  0.01 & 7.41 &   0.04         & 0.09      \\
$\log \varepsilon(\text{O}) $ & $-2.9\times10^{-4}$                     &  0.7                         &  0                         & 0.05  &   0.03  &    0.03        &  0.02 & 8.59 &   0.02         & 0.13      \\
$\log \varepsilon(\text{Ne})$ & $1.1\times10^{-4}$                      & -0.7                         & -0.006                     & 0.04  &   0.04  &    0.04        & 0.01 & 8.39 &  0.06          & 0.11      \\
$\log \varepsilon(\text{Mg})$ & $1.1\times10^{-4}$                      & -1.4                         & -0.028                     & 0.02  &   0.05  &    0.07        & 0.01 & 7.30 &  0.15          & 0.20      \\
$\log \varepsilon(\text{Al})$ & $-4.5\times10^{-5}$                     &  0.1                         & -0.008                     & 0.04  &   0.04  &    0.01        &  0.02 & 5.56 &   0.02         & 0.09      \\
$\log \varepsilon(\text{Si})$ &        0                                &  0.6                         & -0.064                     & 0.1   &    0.01  &    0           &  0.02 & 7.10 &   0.03         & 0.18     \\
$\log \varepsilon(\text{S}) $ & $-1.9\times10^{-4}$                     &  0.6                         & -0.020                     & 0.09  &    0.06  &    0.01        &  0.02 & 6.39 &   0.05         & 0.16     \\
\hline
\end{tabular}
\end{center}
\end{table*}

\subsection{Rotational velocity}
The rotational velocity $v\sin i$ was measured from comparison of
observed and synthetic spectra on 12 non-blended lines (10 --
\ion{O}{ii}, 1 -- \ion{Si}{iii}, 1 -- \ion{Al}{iii}) without
visible distortions. We generated a grid of synthetic spectra for
obtained stellar parameters with various rotational velocities.
The rotational broadening was calculated directly by integrating
intensities produced by {\sc synplot}.

\begin{figure}
\begin{center}
\includegraphics[scale=0.65]{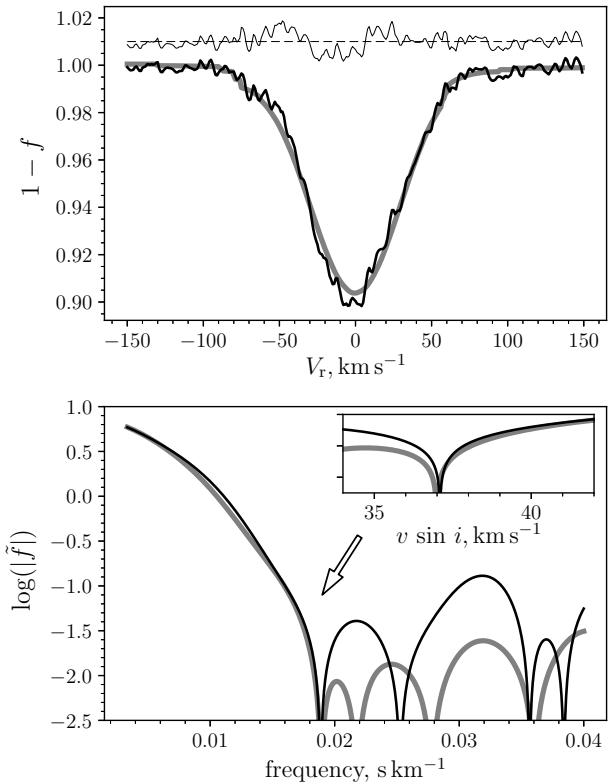}
\end{center}
\caption{The average profiles (the upper panel) and corresponding
Fourier transforms (the lower panel). The black lines are for
observations, the grey lines are for the synthetic spectrum for
$v\sin\,i=37$\,{\kms} and $\sigma_f=22$\,\kms. The thin line on
the upper panel is for the residual between the synthetic and
observed spectra.}\label{fig:vsini}
\end{figure}

Line profiles were transformed into velocity space in the stellar
rest frame using theoretical line positions and average radial
velocity of the star $V_{\rm r}=-124$\,{\kms}. After that all
profiles were interpolated on a single velocity grid and averaged.
Both observed and synthetic profiles were treated in exactly the
same manner. A comparison between the averaged profiles was
carried out in Fourier space \citep{SG76}. The first zero in the
Fourier transform $|\tilde{f}|$ gives $v \sin i$ about 37\,\kms,
and shape of low-frequency part of $|\tilde{f}|$ gives
$\sigma_f=22$\,{\kms} for the Gaussian contribution (see
Fig.\,\ref{fig:vsini}).

To estimate uncertainty of $v\sin i,$ we performed the following
Monte-Carlo calculations. We generated $\sim10^5$ artificial
observations, each of which is a synthetic spectrum for $v\sin
i=37$\,{\kms}, $\sigma_f=22$\,{\kms} with a Gaussian noise with
$\sigma=0.01,$ which is correspond to the noise in the real
observed spectrum. Applying our procedure, we obtain distributions
for $v\sin i$ and $\sigma_f.$ The distribution for $v\sin i$ is
slightly asymmetric, but within $2\sigma$ can be considered as a
Gaussian distribution with $\sigma=6$\,\kms, the distribution of
$\sigma_f$ looks like a Gaussian with $\sigma=1$\,\kms. Removing
the instrumental contribution from $\sigma_f$ reduces it by
1\,\kms. Summarizing, $v\sin i=37\pm6$\,{\kms} and the Gaussian
contribution is $\sigma_f=21\pm1$\,\kms, which can include various terms:
microturbulence, macroturbulence, instrumental profile,
deviations in line positions.

Some asymmetry can be suspected in the observed profile. It
manifests itself in deviations from the synthetic profile at
velocities --50, --25 and +10\,\kms, which makes the blue wing
more steep, then the red one. If this deviations are real, then
they may be signatures of radial expansion of the star or ongoing
convection.

\section{ANALYSIS OF THE EMISSION LINE SPECTRUM}\label{sec5}

Due to the rather low temperature of the central star, the
emission spectrum of the object still contains a very limited set
of lines, which are usually used to diagnose the gas envelope. In
addition, in the absence of a flux calibrated spectrum  it is not
possible to obtain reliable absolute fluxes for the emission
lines. However, using the equivalent widths given in
Table~\ref{emlines} and stellar continuum flux distributions for
the atmospheric parameters defined above, we can obtain reliable
ratios of fluxes in the emission lines.

\subsection{Nebular parameters}

Unfortunately, we cannot estimate plasma parameters from
[\ion{S}{ii}] ratio, since the ratio
$I$($\lambda$6716)/$I$($\lambda$6731)=0.35 indicates that the
electron density is higher than the critical one (of order
$10^4$~cm$^{-3}$). Due to the absence of the $\lambda$5755 line in
the spectrum, we cannot to estimate the electron temperature from
the ratio [\ion{N}{ii}]
($I$($\lambda$6548)+$I$($\lambda$6584))/$I$($\lambda$5755).

Using the PyNeb analysis package \citep{lur15}, we obtained $T_e$
versus $N_e$ contours for the observed [\ion{N}{I}]
$I$($\lambda$5198)/$I$($\lambda$5200) and [\ion{O}{I}]
($I$($\lambda$6300)+$I$($\lambda$6363))/$I$($\lambda$5577)
diagnostic ratios of 1.9 and 8.8, respectively. In the $T_e$
range from 5~000 to 20~000~K and $\log N_e (\rm cm^{-3})$ from 3 to 8, the
curves have no intersection. The nebular line ratios for
[\ion{N}{i}] indicates that these lines are formed in a partial
ionized region of $N_e\sim3~000$~cm$^{-3}$, whereas  [\ion{O}{i}]
nebular/auroral ratio suggest  $N_e > 10^6$ cm$^{-3}$.
\citet{bautista99} investigate the effects of photoexcitation of
[\ion{N}{i}] and [\ion{O}{i}] lines by stellar continuum radiation
under nebular conditions and found that the [\ion{N}{i}] optical
lines at 5198 \AA\ and 5200 \AA\ are affected by fluorescence in
many objects. In the presence of radiation fields there is no
unique solution in terms of $N_e$ and $T_e$ for an observed
[\ion{N}{i}] spectrum.

The comparison of the theoretical from \citet{bpp96} and the
observed line ratio $I$($\lambda$7412)/$I$($\lambda$7379)=0.31
testifies the pure collisional excitation of [\ion{Ni}{II}] lines
in the gaseous shell of IRAS 18379--1707 with $N_e > 10^6$
cm$^{-3}$. This result is consistent with conclusion \citet{bpp96}
about strong intercorrelations between [\ion{Ni}{II}] and
[\ion{O}{I}] emission in gaseous nebulae, which suggests that they
stem from coincident zones.

\subsection{Expansion velocities}

The expansion velocities of the nebula calculated from the
formula:
$V_{\text{exp}}=1/2(V_\text{FWHM}^2-V_\text{instr}^2)^{1/2}$,
where $V_\text{FWHM}$ is the velocity corresponding to the full
width at half maximum (FWHM) and $V_\text{instr}$ (6 \kms) is the
instrumental broadening. The adopted $V_{\text{exp}}$ for each ion
are given in Table~\ref{tab:vel}. [\ion{N}{II}](1F) $\lambda$6584
line is blended with the absorption component of the
\ion{C}{II}(2) $\lambda$6583 and has not been used to estimate the
expansion velocity.

\begin{figure}
\includegraphics[width=\columnwidth]{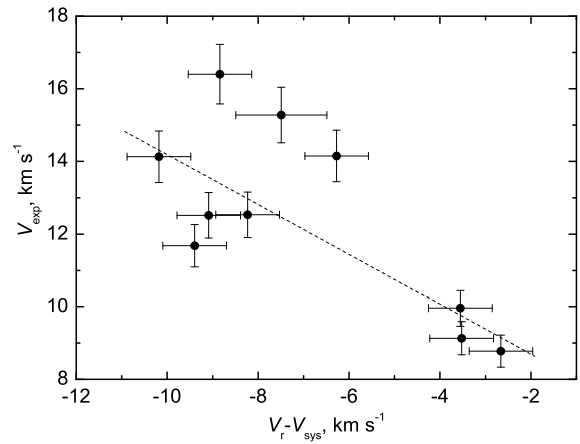}
\caption{Expansion velocity vs. $V_{\text{r}}-V_{\text{sys}}$ for
different ions with a linear trend of the data (dashed line).}
\label{Vexp}
\end{figure}

In the case of a spherical symmetric expanding envelope, spectral
lines should possess the same radial velocity as the central star.
Nevertheless, the emission lines are blue-shifted relative to the
stellar spectrum, moreover this shift is greater for greater
values of $V_{\rm exp}$, see Fig.\,\ref{Vexp}. Such behaviour can
be explained if the observer for some reason (for example, due to
an intrinsic absorption in the envelope) receives less light from
the back parts of the envelope than from the front ones, which
leads to the appearance of line asymmetry and the dependence of
the line shift on the expansion velocity $V_{\rm exp}$.

The expansion velocities given in Table~\ref{tab:vel} shows that
the low excitation nebula present around the star is slowly
expanding. However at present the nebula is very compact and it is
not resolved. As the star evolves to higher $T_{\rm eff}$ value it
will photoionize the nebula and there will be hot and fast stellar
wind from the star, by then the present compact nebula will expand
and grow in size. Once we can measure the angular size of the
nebula we can calculate the linear size as the distance to the
star is known. Using the linear size of the nebula and expansion
velocity we can calculate the age of the nebula (for example
please see \citet{part93, part95} in the case of SAO 244567).

\begin{table}
\caption{Expansion velocities.}
 \label{tab:vel}
 \begin{center}
\begin{tabular}{lccc}
\hline
Identification    &$\lambda_\text{lab.}$ & $V_{\text{FWHM}}$ & $V_{\text{exp}}$ \\
                  & \AA                & {\kms}              & {\kms} \\
\hline
[\ion{Fe}{II}](7F) & 4287.39 & 24.12 & 11.68\\

[\ion{N}{I}](1F)   & 5197.90 & 31.14 & 15.28\\

[\ion{N}{I}](1F)   & 5200.26 & 25.77 & 12.53\\

[\ion{O}{I}](1F)   & 6300.30 & 20.79 & 9.96\\

[\ion{O}{I}](1F)   & 6363.78 & 19.22 & 9.13\\

[\ion{N}{II}](1F)  & 6548.05 & 28.89 & 14.13\\

[\ion{S}{II}](2F)  & 6730.82 & 25.74 & 12.52\\

[\ion{Ni}{II}](2F) & 7377.83 & 28.93 & 14.15\\

[\ion{Cr}{II}](3F) & 8000.07 & 33.35 & 16.40\\

[\ion{C}{I}](3F)   & 8727.13 & 18.55 & 8.77\\

\hline
\end{tabular}
\end{center}
\end{table}

\section{DISCUSSION}\label{sec6}

Based on high-resolution ($R\sim48~000$) observations we have
studied the optical spectrum of the early B-supergiant with IR
excess IRAS~18379--1707. At wavelengths from 3700 to 8820 \AA,
numerous absorption and emission lines have been identified, their
equivalent widths and corresponding radial velocities have been
measured. Using non-LTE model atmospheres, we have obtained the
effective temperature $T_{\text{eff}}=18~000\pm1000$ K, gravity
$\log g=2.25\pm0.08$, microturbulence velocity $\xi_{\rm
t}=10\pm4$\,\kms and rotational velocity $v\sin i=37\pm6$\,{\kms}.
The temperature agrees within error limits with the previously
determined value from \citet{gp04} and is lower than that
of \citet{c09} (see Table~\ref{param}). The parameters $T_{\rm
eff}$ and $\log g$ lead from the \citet{strai82} calibration to
the spectral type B2-B3I  which consistent with B2.5Ia from
\citet{v98}.

\subsection{Abundance}\label{abund}

As stressed by \citet{stas06} after, e.g.,  \citet{ml92}, \ion{Fe}{},
cannot be used for post-AGB stars as  the metallicity indicator in stellar
atmospheres because of possible strong depletion in dust grains in
a former stage and subsequent ejection of the grains. On the other
hand, \ion{S}{} does not get depleted even if there is a dust-gas
separation, because \ion{S}{} does not get condensed into dust
grains and \ion{S}{} trace the original metallicity of the star.
For IRAS 18379--1707 [\ion{S}{}/\ion{H}{}]=--0.73, hence we
conclude that the star is metal poor. \ion{Al}{} and \ion{Si}{}
show deficiency by --0.92, and --0.43 dex, respectively.

One has to note, that in IRAS 18379--1707 the uncertainty on the
carbon abundance prevents an accurate C/O number ratio for this
star. However, if we accept that the carbon abundance derived from
the $\lambda$3919, 3921 and 4267 lines are closer to the truth
than that measured from the $\lambda$6578 line, which is
associated with a large equivalent width and seems to have its
origin in the outflow, then the carbon abundance is $\log
\varepsilon(\rm C)=7.76$ and C/O<1.

\citet{stas06} offered to identify objects that have experienced
third dredge-up as those objects in which (C+N+O)/S is larger than
in the Sun. The estimates of the CNO abundances in
IRAS~18379--1707 indicate that these elements are overabundant
with [(C+N+O)/S]$=+0.5\pm0.2$ suggesting that the products of
helium burning have been brought to the surface as a result of
third dredge-up on the AGB.

\subsection{Mass and luminosity}\label{mass}

Our analysis of the high-resolution optical spectrum of
IRAS~18379--1707 with other published results (infrared colours
similar to PNe, presence of a circumstellar envelope) confirms
that the star is indeed in the post-AGB phase.

To determine the stellar mass, we compared $T_{\text{eff}}$ and
$\log g$ to stellar evolutionary calculations for H-rich post-AGB
stars that have recently been presented by \citet{bert16}.
Unfortunately, the models for metallicity of $Z$=0.003 are not yet
calculated, so we used two grids with initial metallicities of
$Z$=0.010 and $Z$=0.001.

\begin{figure}
\includegraphics[width=\columnwidth]{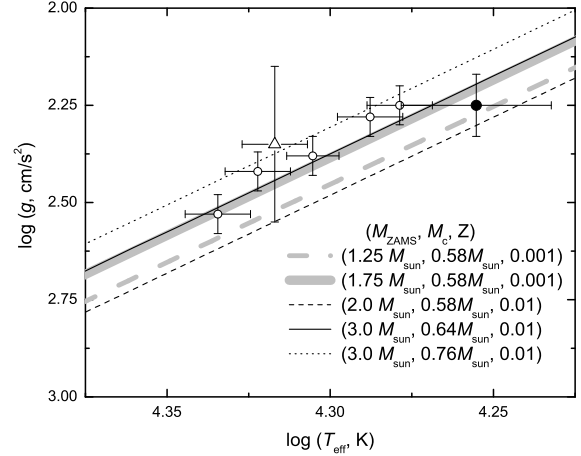}
\caption{Location of IRAS~18379--1707 (filled circle) in the
$\log T_\text{eff}-\log g$ plane with $Z$ = 0.001 and $Z$ = 0.01
single-star evolutionary tracks of \citet{bert16}. Also shown are
the positions of other hot post-AGB stars: V886~Her
\citep{ryans03} (open triangle) and five objects from
\citet{mello12} (open circles).} \label{models}
\end{figure}

Fig.~\ref{models} shows the evolutionary tracks of \citet{bert16}
for  metallicities of $Z$ = 0.01 and $Z$ = 0.001 plotted in the
$\log T_\text{eff}-\log g$ diagram. From the derived atmospheric
parameters we find IRAS~18379--1707 to be located between the two
mass tracks for  $Z$ = 0.01, implying a current mass $M_\text{c}$
of 0.58 to 0.64~M$_{\odot}$ and a initial mass of the progenitor
$M_{\text{ZAMS}}$ of 2.0 to 3.0~M$_{\odot}$. The grid with initial
metallicity of $Z$ = 0.001 yield a current mass of
$M_\text{c}=0.58~M_{\odot}$ and a initial mass of 1.25 to
1.75~M$_{\odot}$. So we estimate the core mass $M_\text{c}\sim
0.58M_{\odot}$ and the mass of the progenitor $M_\text{ZAMS}\sim
1.5M_{\odot}$.

As seen from Fig.~\ref{models} that IRAS~18379--1707 is placed in
a region of the  $\log T_{\text{eff}}$ and $\log g$ diagram where
other hot post-AGB objects have been observed, viz. V886 Her
\citep{ryans03}, TYC 6234-178-1, LSS 4634, LS 3099, LS IV-12$^{\circ}$111,
LSE 63 \citep{mello12}. Here we compare the position of post-AGB
stars in the  $\log T_{\text{eff}}$--$\log g$  diagram, for which
non-LTE analysis has been performed. As noted
by \citet{ryans03} the LTE analyses yielded effective temperature
estimates and  the gravities significantly higher than those from
the non-LTE analyses.

With the core mass $M_{\text{c}}\sim 0.58M_{\odot}$ and
$T_{\text{eff}}=18~000$~K, the star falls on the horizontal part
of the post-AGB evolutionary track on the HR diagram with
$\log(L/L_{\odot})\sim 3.95$.

\subsection{Distance and location in Galaxy}

IRAS~18379-1707 is present in the \textit{Gaia} data release DR2
\citep{gaia18} with the parallax $\pi=0.2593\pm0.0648$  mas and
proper motion $\mu_{\alpha}=-2.065\pm0.089$ mas/yr,
$\mu_{\delta}=0.410\pm0.075$ mas/yr. \textit{Gaia} DR2 parallaxes
have a zero-point error that is different for different objects
\citet{arenou18}, but for weak stars it averages --0.05 mas, in
the direction of increasing parallaxes and, accordingly,
decreasing distances. If we add 0.05 to parallax 0.2593, then we
get 0.3093 mas and the distance to the object decreases to
$d=3.24^{+0.86}_{-0.57}$ kpc.

The star is located at the Galactocentic distance $r_g\sim4.5$ kpc
with the azimuthal angle $\theta=11.^{\circ}8$ between the
Sun--Galactic center line and the direction to the star and at the
distance from the Galactic plane of $z=-306_{-81}^{+54}$ pc. Its
velocity components calculated in the direction of the Galactic
radius vector ($U$), azumuthal direction ($V$), and perpenducular
to the Galactic plane ($W$) are  147, 171, 50 \kms, respectively.
The star rotates slower than it should due to the Galactic
rotation curve by 58 \kms ($V_{res}=-58$ \kms). The total velocity
with respect to the Galactic center is 230 km/s. Generally the
velocity deviations from the circular velocity can be due to
perturbation from the Galacic bar \citep{melnik19}.

Among post-AGB supergiants there are few high velocity stars.
Among the cooler post-AGB supergiants HD 56126 (+105 \kms)
\citep{part92}, HD~179821 (+81.8 \kms) \citep{part19}, were found
and among hotter post-AGB stars LS~III +52$^{\circ}$24
(IRAS~22023+5249) (--148 \kms) \citep{sarkar12}, and
BD+33$^{\circ}$2642 (--100 \kms) \citep{napiw94}.

Using the determined luminosity from Sect.~\ref{mass}, $V=11.93$
mag \citep{r98}, $E(B-V)=0.71$ mag from \citet{gp04}, $R_V=3.1$
and the bolometric corrections ($BC$) in the calibration of
\citet{vacca96} we derived the distance $d\sim3.8$~kpc from $\log
d=V+5-3.1E(B-V)-M_{bol\odot}-BC+2.5\log(L/L_\odot)$. This value is
close to that derived from the \textit{Gaia} parallax measurement.
This result also confirms the conclusion that IRAS 18379--1707 is
a low-mass post-AGB star, and not a massive Population I B
supergiant with a typical luminosity of about
$\log(L/L_{\odot})\sim5.0$.

\subsection{Photometric variability}

IRAS~18379--1707 is suspected of variability and in the General
Catalog of Variable Stars \citep{samus17} is designated as NSV
24542, but the type of variability has not yet been determined.

\begin{figure}
 \includegraphics[width=\columnwidth]{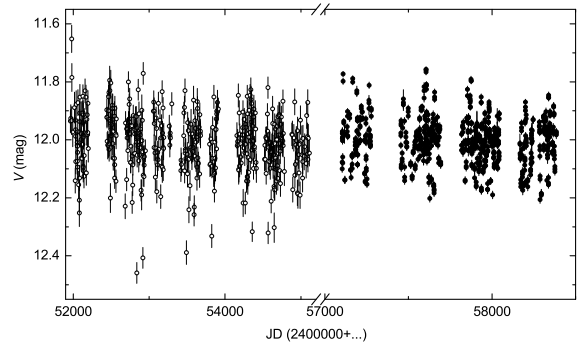}
 \caption{ASAS-3 (the open circles) and ASAS-SN (the filled  circles)
  light curve of IRAS~18379--1707 from 2009 to 2018.}
 \label{lc}
\end{figure}

To analyse the photometric variability, we used two sources of
data: All Sky Automated Survey (ASAS-3, \citet{pojm02}) and All
Sky Automated Survey for Super-Novae (ASAS-SN, \citet{shap14,
koch17}). IRAS 18379--1707 was monitored with the ASAS-3 in $V$
photometric band since  February 22, 2001 up to October 6, 2009.
We used the good quality (symbol A) measurements made with
aperture 1 (15$\arcsec$). ASAS-SN $V$-band data span a time
interval from March 21, 2015  to September 23, 2018. The ASAS-3
and ASAS-SN data points are shown in Fig.~\ref{lc} with their
corresponding errorbars. The mean brightness in $V$ band with the
standard deviation ($SD$), the number of observations (N) and the
mean accuracy of the measurements ($\sigma V$) from ASAS and
ASAS-SN are listed in Table~\ref{tab:asas}.

The star displays a brightness variation on a scale of a few
nights with a maximum amplitude (peak-to-peak) of up to 0.2 mag in
the $V$ band. We performed a period analysis with the
period-finding program Period04 \citep{lb05} and we did not find
any reliable period to describe the variations. The pattern of
variability for IRAS 18379--1707 is very similar to the
photometric behaviour of the other hot post-AGB, early B
supergiants with IR excesses \citep{a07, a13, a14, a18}. All of
them display fast irregular photometric variability with
amplitudes from 0.2 to 0.4 in the $V$-band. For a number of the
hot post-AGB objects, spectral variability was also detected,
which is expressed in line variations in both shape and intensity
\citep{parth93, gl97b, a12, kcpsy14}. These variations, as well as
photometric variability, occur on a scale of a few days. In
addition, IRAS 18379--1707, as well as other hot post-AGB objects
\citep{sarkar05, mello12, ryans03}, shows the H$\alpha$
P-Cygni profile ongoing mass-loss. Therefore, most likely, the
photometric variability is associated with a variable stellar
wind. However, other causes of variability are not excluded, for
example, short-period pulsations for the detection of which
observations with better temporal resolution than those of ASAS
and ASAS-SN are necessary.

\begin{table}
\caption{Summary of photometric observations of IRAS~18379--1707}
 \label{tab:asas}
 \begin{center}
\begin{tabular}{lcccc}
\hline
Source  &   $V$, mag& $SD$, mag &      N & $\sigma V$, mag\\
\hline
ASAS-3  &     12.01 &   0.11    &     460&   0.041\\
ASAS-SN &     12.00 &   0.08    &     720&   0.013\\

\hline
\end{tabular}
\end{center}
\end{table}

\section{CONCLUSIONS}\label{sec7}

From a non-LTE analysis of high resolution spectrum of hot
post-AGB candidate IRAS~18379--1707 (LS 5112) we find its
$T_{\text{eff}}=18000$ K, $\log g=2.25$ and find that it is a
metal-poor [M/H]=--0.6 high velocity star $V_r=-124$ \kms. Oxygen,
Carbon and Nitrogen are slightly overabundant relative to Sulfur
suggesting that the products of helium burning have been brought
to the surface as a result of third dredge-up on the AGB.

We found permitted and forbidden emission lines of several
elements in its spectrum. The nebular emission lines indicates the
presence of a low excitation nebula around this post-AGB star in
agreement with other studies mentioned in the paper (see for
example \citet{gf15} who found evidence for the presence of
bipolar flow). The presence of several P-Cygni lines in the
spectrum indicates ongoing post-AGB mass-loss. The mean radial
velocity as measured from emission features of the envelope is
$V_\text{r}=-130.9\pm0.5$ {\kms}.

We measured the radial velocities from the Doppler shifts of many
spectral lines and discovered that LS 5112 is a high velocity
star.

From the \textit{Gaia} DR2 parallax we find the distance to be 3.2
kpc. From the derived $T_{\text{eff}}$, $\log g$, and luminosity
and placing it on the recent post-AGB evolutionary tracks we
conclude that it is a post-AGB star of core mass about
0.58$M_{\odot}$.

\section{ACKNOWLEDGMENTS}

This research has made use of the SIMBAD database, operated at
CDS, Strasbourg, France, and SAO/NASA Astrophysics Data System. AD
acknowledges the support from the Program of development of M.V.
Lomonosov Moscow State University (Leading Scientific School
{\lq}Physics of stars, relativistic objects and galaxies{\rq}). We
would like to thank Dr. A.M. Melnik for her help in calculations
related to the kinematics of the star. We are also grateful to the
anonymous referee for the careful reading of the manuscript and
the numerous important remarks that helped to improve the paper.

\appendix
\newpage
\onecolumn
\section{Emission lines in the spectrum of IRAS~18379--1707}

\begin{center}
\begin{longtable}{ccccc}
  \caption{Emission lines in the spectrum of IRAS~18379--1707.}
\label{emlines}\\

\hline

   $\lambda_{\text{obs.}}$& $\lambda_\text{lab.}$ & Identification &  $EW$  & $V_\text{r}$\\
             \AA\         &      \AA\             &                &   \AA\ &      {\kms} \\

  \hline

  \endfirsthead

   \multicolumn{2}{l}{continued Table\ref{emlines}}\\

   \hline
   $\lambda_\text{obs.}$ & $\lambda_\text{lab.}$ & Identification &  $EW$  & $V_\text{r}$\\
             \AA\        &      \AA\             &                &   \AA\ &      {\kms} \\

   \hline
   \endhead

3854.42& 3856.02& \ion{Si}{II}(1)    & 0.072& -124.47\\
3860.98& 3862.59& \ion{Si}{II}(1)     & 0.043& -124.81\\
4199.31& 4201.17& [\ion{Ni}{II}]     & 0.007& -132.93\\
4242.15& 4243.98& [\ion{Fe}{II}](21F)& 0.024& -129.01\\
4274.96& 4276.83& [\ion{Fe}{II}]     & 0.030& -131.25\\
4285.49& 4287.39& [\ion{Fe}{II}](7F) & 0.092& -133.40\\
4350.85& 4352.78& [\ion{Fe}{II}](21F)& 0.015& -132.97\\
4357.39& 4359.34& [\ion{Fe}{II}](7F) & 0.069& -133.83\\
4366.31& 4368.13& \ion{O}{I}(5)      & 0.019& -124.43\\
4411.83& 4413.78& [\ion{Fe}{II}](6F) & 0.054& -132.74\\
4414.31& 4416.27& [\ion{Fe}{II}](6F) & 0.041& -133.13\\
4450.08& 4452.11& [\ion{Fe}{II}](7F) & 0.026& -136.69\\
4455.96& 4457.95& [\ion{Fe}{II}](6F) & 0.029& -133.54\\
4472.94& 4474.91& [\ion{Fe}{II}](7F) & 0.015& -131.72\\
4772.57& 4774.74& [\ion{Fe}{II}](20F)& 0.014& -136.17\\
4812.41& 4814.55& [\ion{Fe}{II}](20F)& 0.055& -132.98\\
4887.44& 4889.63& [\ion{Fe}{II}](4F) & 0.014& -134.18\\
4903.19& 4905.35& [\ion{Fe}{II}](20F)& 0.021& -131.95\\
5038.90& 5041.03& \ion{Si}{II}(5)     & 0.049& -126.67\\
5053.90& 5055.98& \ion{Si}{II}(5)     & 0.122& -123.24\\
5109.38& 5111.63& [\ion{Fe}{II}](19F)& 0.017& -131.91\\
5125.29& 5127.35& \ion{Fe}{III}(5)    & 0.017& -120.21\\
5153.94& 5156.12& \ion{Fe}{III}(5)    & 0.021& -127.45\\
5156.50& 5158.81& [\ion{Fe}{II}](19F)& 0.064& -134.15\\
5195.62& 5197.90& [\ion{N}{I}](1F)   & 0.062& -131.49\\
5197.96& 5200.26& [\ion{N}{I}](1F)   & 0.033& -132.23\\
5259.30& 5261.61& [\ion{Fe}{II}](19F)& 0.048& -131.50\\
5271.02& 5273.35& [\ion{Fe}{II}](18F)& 0.023& -132.46\\
5296.69& 5298.97& \ion{O}{I}(26)     & 0.020& -128.66\\
5331.29& 5333.65& [\ion{Fe}{II}](19F)& 0.014& -132.66\\
5510.27& 5512.77& \ion{O}{I}(25)     & 0.018& -135.96\\
5552.51& 5554.95& \ion{O}{I}(24)     & 0.023& -131.71\\
5574.94& 5577.34& [\ion{O}{I}](3F)   & 0.021& -128.92\\
5887.41& 5889.95& \ion{Na}{I}(1)       & 0.068& -129.30\\
5893.35& 5895.92& \ion{Na}{I}(1)    & 0.020& -130.82\\
5955.16& 5957.56& \ion{Si}{II}(4)     & 0.153& -120.95\\
5976.39& 5978.93& \ion{Si}{II}(4)     & 0.258& -127.37\\
6043.71& 6046.38& \ion{O}{I}(22)     & 0.057& -132.12\\
6297.62& 6300.30& [\ion{O}{I}](1F)   & 0.220& -127.55\\
6344.41& 6347.11& \ion{Si}{II}(2)     & 0.138& -127.53\\
6361.07& 6363.78& [\ion{O}{I}](1F    & 0.063& -127.52\\
6368.65& 6371.37& \ion{Si}{II}(2)     & 0.050& -127.93\\
6545.12& 6548.05& [\ion{N}{II}](1F)      & 0.075& -134.18\\
6580.47& 6583.45& [\ion{N}{II}](1F)      & 0.222   & -135.85\\
6663.90& 6666.80& [\ion{Ni}{II}](2F) & 0.042& -130.63\\
6713.50& 6716.44& [\ion{S}{II}](2F)      & 0.020& -131.05\\
6727.83& 6730.82& [\ion{S}{II}](2F)      & 0.057& -133.09\\
6999.05& 7002.13& \ion{O}{I}(21)     & 0.055& -131.71\\
7152.01& 7155.17& [\ion{Fe}{II}]     & 0.022& -132.58\\
7228.40& 7231.33& \ion{C}{II}(3)     & 0.146& -121.32\\
7233.58& 7236.42& \ion{C}{II}(3)     & 0.241& -117.49\\
7251.15& 7254.36& \ion{O}{I}(20)     & 0.078& -132.61\\
7374.62& 7377.83& [\ion{Ni}{II}](2F) & 0.312& -130.27\\
7408.39& 7411.61& [\ion{Ni}{II}](2F) & 0.099& -130.30\\
7420.40& 7423.64& \ion{N}{I}(3)      & 0.017& -130.72\\
7464.96& 7468.31& \ion{N}{I}(3)      & 0.071& -134.69\\
7509.92& 7513.08& \ion{Fe}{II}(J)    & 0.046& -126.09\\
7873.78& 7877.05& \ion{Mg}{II}(8)    & 0.144& -124.45\\
7893.01& 7896.37& \ion{Mg}{II}(8)    & 0.296& -127.56\\
7996.50& 8000.07& [\ion{Cr}{II}](1F) & 0.108& -133.60\\
8121.77& 8125.30& [\ion{Cr}{II}](1F) & 0.090& -130.18\\
8231.18& 8234.64& \ion{Mg}{II}(7)    & 0.156& -125.96\\
8238.76& 8242.34& \ion{N}{I}(7)      & 0.083& -130.21\\
8297.14& 8300.99& [\ion{Ni}{II}](2F) & 0.065& -138.90\\
8304.71& 8308.51& [\ion{Cr}{II}](1F) & 0.048& -137.07\\
8442.76& 8446.25& \ion{O}{I}(3)      & 2.452& -124.00\\
8613.11& 8616.96& [\ion{Fe}{II}](13F)& 0.071& -134.02\\
8679.50& 8683.40& \ion{N}{I}(1)      & 0.072& -134.82\\
8699.50& 8703.25& \ion{N}{I}(1)      & 0.108& -129.12\\
8707.89& 8711.70& \ion{N}{I}(1)      & 0.068& -131.25\\
8723.44& 8727.13& [\ion{C}{I}](3F)   & 0.095& -126.66\\
9114.36& 9218.25& \ion{Mg}{II}(1)    & 1.310& -126.51\\

  \hline

\end{longtable}
\end{center}

\section{ Radial velocities of the stellar absorption lines}
\begin{table}
\caption{Absorption lines used for measurement of the stellar
radial velocity.}
 \label{tab:abslines}
 \begin{center}
\begin{tabular}{lccccc}
 \hline
  ion & $\lambda_{\rm lab}$ & $E_{l}$  &  $EW$              &  $V_{\rm r}$ &  $\sigma_{V_{\rm r}}$\\
      &       \AA\          &   eV     & m\AA\             &   {\kms}      &   {\kms}     \\
  \hline
\ion{He}{i}        & 3867.49  & 21.0   &  120      &  -130.3      &  2.6 \\
\ion{He}{i}        & 3871.78  & 21.2   &   96      &  -121.5      &  2.7 \\
\ion{O}{ii}        & 3911.97  & 25.7   &   77      &  -123.8      &  1.5 \\
\ion{C}{ii}$^b$    & 3919.14  & 16.3   &  105      &  -129.4      &  2.0 \\
\ion{C}{ii}$^b$    & 3920.61  & 16.3   &  100      &  -125.7      &  1.5 \\
\ion{He}{i}        & 3926.53  & 21.2   &  169      &  -122.5      &  1.2 \\
\ion{O}{ii}        & 3945.04  & 23.4   &   53      &  -128.0      &  2.7 \\
\ion{O}{ii}        & 3973.21  & 23.4   &   82      &  -127.8      &  1.3 \\
\ion{N}{ii}        & 3995.00  & 18.5   &  114      &  -124.9      &  1.1 \\
\ion{O}{ii}        & 4069.78  & 25.6   &  140      &  -125.1      &  1.1 \\
\ion{O}{ii}        & 4072.14  & 25.6   &  112      &  -123.8      &  1.3 \\
\ion{O}{ii}        & 4075.85  & 25.7   &  117      &  -125.6      &  0.8 \\
\ion{O}{ii}        & 4153.29  & 25.8   &   72      &  -126.7      &  1.7 \\
\ion{He}{i}        & 4169.05  & 21.2   &   66      &  -124.8      &  1.4 \\
\ion{O}{ii}        & 4185.44  & 28.4   &   56      &  -128.5      &  1.9 \\
\ion{O}{ii}        & 4189.78  & 28.4   &   64      &  -129.3      &  1.4 \\
\ion{S}{iii}       & 4253.61  & 18.2   &   88      &  -124.0      &  1.3 \\
\ion{O}{ii}        & 4317.15  & 23.0   &  106      &  -126.3      &  0.9 \\
\ion{O}{ii}        & 4319.63  & 23.0   &  103      &  -124.9      &  0.8 \\
\ion{O}{ii}        & 4345.52  & 23.0   &  105      &  -123.6      &  1.0 \\
\ion{O}{ii}        & 4347.44  & 25.7   &   53      &  -124.7      &  1.9 \\
\ion{O}{ii}        & 4349.40  & 23.0   &  153      &  -121.1      &  0.6 \\
\ion{O}{ii}        & 4351.27  & 25.7   &   58      &  -126.4      &  0.8 \\
\ion{O}{ii}        & 4366.89  & 23.0   &  100      &  -127.3      &  1.0 \\
\ion{O}{ii}        & 4414.90  & 23.4   &  122      &  -123.6      &  0.8 \\
\ion{He}{i}        & 4437.54  & 21.2   &   75      &  -126.1      &  1.2 \\
\ion{Si}{iii}      & 4552.62  & 19.0   &  225      &  -120.9      &  0.5 \\
\ion{Si}{iii}      & 4567.84  & 19.0   &  187      &  -122.8      &  0.6 \\
\ion{Si}{iii}      & 4574.75  & 19.0   &  121      &  -125.3      &  0.7 \\
\ion{O}{ii}        & 4590.97  & 25.7   &   93      &  -128.2      &  0.8 \\
\ion{O}{ii}        & 4596.16  & 25.7   &   89      &  -126.8      &  1.0 \\
\ion{O}{ii}        & 4638.85  & 23.0   &   95      &  -124.4      &  0.8 \\
\ion{O}{ii}        & 4641.81  & 23.0   &  130      &  -123.1      &  0.7 \\
\ion{O}{ii}        & 4649.14  & 23.0   &  183      &  -118.7      &  0.6 \\
\ion{O}{ii}        & 4650.80  & 23.0   &  106      &  -124.5      &  0.9 \\
\ion{O}{ii}        & 4661.63  & 23.0   &  101      &  -125.0      &  0.6 \\
\ion{O}{ii}        & 4676.24  & 23.0   &  100      &  -124.0      &  0.8 \\
\ion{O}{ii}        & 4699.16  & 28.5   &   62      &  -126.2      &  1.1 \\
\ion{O}{ii}        & 4705.34  & 26.2   &   70      &  -125.3      &  1.0 \\
\ion{O}{ii}        & 4710.01  & 26.2   &   38      &  -129.4      &  1.8 \\
\ion{Si}{iii}      & 4819.71  & 26.0   &   54      &  -128.6      &  1.6 \\
\ion{Si}{iii}      & 4828.96  & 26.0   &   43      &  -125.6      &  2.0 \\
\ion{O}{ii}        & 4924.47  & 26.3   &   73      &  -131.3      &  1.3 \\
\ion{N}{ii}        & 5001.37  & 20.6   &   81      &  -127.8      &  1.1 \\
\ion{N}{ii}        & 5005.15  & 20.7   &   56      &  -128.8      &  1.9 \\
\ion{N}{ii}        & 5666.62  & 18.5   &   65      &  -123.8      &  1.4 \\
\ion{N}{ii}        & 5676.02  & 18.5   &   71      &  -123.4      &  1.3 \\
\ion{N}{ii}        & 5679.56  & 18.5   &  135      &  -122.2      &  1.0 \\
\ion{Al}{iii}      & 5696.57  & 15.6   &   95      &  -123.7      &  0.9 \\
\ion{Al}{iii}      & 5722.71  & 15.6   &   55      &  -130.5      &  1.8 \\
\ion{Si}{iii}      & 5739.74  & 19.7   &  161      &  -119.9      &  0.5 \\
\hline
\end{tabular}
\end{center}
To more accurate treatment of blended/multicomponent lines, we
measure $\lambda_{\rm lab}$ using the theoretical spectrum
generated by {\sc synplot}. $^b$ marks probably blended lines.
\end{table}

\begin{table}
\caption{Absorption lines with extended blue wings.}
 \label{tab:blueshifted}
 \begin{center}
\begin{tabular}{lccccc}
 \hline
  ion & $\lambda_{\rm lab}$ & $E_{l}$  & $EW$                &  $V_{\rm r}$ &  $\sigma_{V_{\rm r}}$\\
      &       \AA\          &   eV     & m\AA\             &   {\kms}      &   {\kms}     \\
  \hline
\ion{Si}{ii}    &   4128.07  &  9.8   &  113      &  -141.7      &  1.5 \\
\ion{Si}{ii}    &   4130.90  &  9.8   &  123      &  -145.1      &  1.5 \\
\ion{Mg}{ii}    &   4481.20  &  8.9   &  211      &  -137.3      &  1.0 \\
\ion{S}{ii}     &   5640.01  & 14.1   &  142      &  -129.3      &  0.9 \\
\ion{S}{ii}     &   5647.01  & 13.7   &   91      &  -140.4      &  1.6 \\
\ion{S}{ii}     &   5659.79  & 13.7   &   49      &  -125.1      &  2.3 \\
\ion{Ne}{i}     &   5852.49  & 16.8   &   77      &  -144.2      &  1.1 \\
\ion{S}{ii}     &   6312.75  & 14.2   &   75      &  -147.0      &  2.8 \\
\ion{Ne}{i}     &   6334.43  & 16.6   &   46      &  -134.8      &  1.9 \\
\ion{Ne}{i}     &   6402.25  & 16.6   &  177      &  -128.7      &  0.7 \\
\ion{Ne}{i}     &   7032.41  & 16.6   &   71      &  -137.3      &  1.6 \\
  \hline
\end{tabular}
\end{center}
\end{table}

\end{document}